\documentclass[preprint,amsmath,trackchanges,longbib]{aastex701}

\usepackage{cureport}

\newcommand{\mps}{~\text{m~s$^{-1}$}}
\newcommand{\kmps}{~\text{km~s$^{-1}$}}
\newcommand{\cmps}{~\text{cm~s$^{-1}$}}

\newcommand{\M}{\mathcal{M}}

\newcommand{\wem}{\lambda_0}  % emitter rest-frame wavelength (bare)
\newcommand{\tem}{t_0}  % emitter rest-frame time (bare)
\newcommand{\aspecem}{F_0}  % adjusted spectrum in emitter rest frame
\newcommand{\daspecem}{F'_0}  % adjusted spectrum wavelength derivative

\newcommand{\wob}{\lambda}  % observer-frame wavelength
\newcommand{\tob}{t}  % observer-frame time
\newcommand{\pflux}{\Phi}  % Photon number flux
\newcommand{\aspec}{F}  % adjusted spectrum

\newcommand{\vlos}{v_{\rm los}}  % LOS velocity
\newcommand{\zz}{z}  % redshift
\newcommand{\dop}{\mathcal{D}}  % Doppler factor

\newcommand{\spar}{\mathcal{S}}  % spectral shape parameters (for SDF)
\newcommand{\trans}{^\mathsf{T}}  % matrix transpose

\newcommand{\dspec}{F}  % dynamic spectrum
\newcommand{\cdspec}{\tilde{F}}  % centered dynamic spectrum
\newcommand{\sbf}{\psi}  % spectral basis function (speclet)
\newcommand{\tbf}{\chi}  % temporal basis function (modulator)

\newcommand{\serr}{\epsilon}  % spectrum error (noise)
\let\proglang=\textsf

\newcommand{\matlab}{\proglang{MATLAB}}

\let\swpack=\texttt
\newcommand{\chebfun}{\swpack{Chebfun}}

\graphicspath{{figures/}}

\begin{document}

\title{Modeling time series of high-resolution spectra\\ using the asymmetric Hilbert-Schmidt decomposition\\ with application to radial velocity exoplanet searches}

\author[orcid=0000-0000-0000-0001,gname=Kevin,sname=Tao]{Kevin Tao}
\affiliation{Department of Statistics and Data Science, Cornell University}
\email[show]{kt575@cornell.edu}  

\author[orcid=0000-0003-4692-4607,gname=Thomas, sname='Loredo']{Thomas Loredo} 
\affiliation{Cornell Center for Astrophysics and Planetary Science (CCAPS), Cornell University}
\affiliation{Department of Statistics and Data Science, Cornell University}
\email[show]{thomas.loredo@cornell.edu}
\correspondingauthor{Thomas Loredo}

\author[orcid=0000-0002-6713-2257,gname=David,sname=Ruppert]{David Ruppert}
\affiliation{Department of Statistics and Data Science, Cornell University}
\email[show]{dr24@cornell.edu}

% \collaboration{all}{The Terra Mater collaboration}

\begin{abstract}

We describe a framework for building low-rank separable expansions to model  spectroscopic data from observations of systems with time-varying spectra, in settings where the spectral dimension is finely sampled (in wavelength), but the temporal dimension is more sparsely sampled, possibly with very irregular sampling.
The motivating application is detection and characterization of small exoplanets, particularly Earth-like planets around Sun-like stars.
Such planets cannot be seen directly due to the glare of the host star, but can be detected via the minute Doppler shifts of the host star spectrum due to the star's motion as the planet and star revolve around their mutual center of mass.
Current instruments are reaching the instrumental precision necessary to detect the $\approx 10\,$\cmps\ motions corresponding to an exo-Earth orbiting a Sun-like star.
But the small Doppler shifts from such low-amplitude orbital motion can be masked or mimicked by the effects of stellar activity on the surface of the host star---starspots (dark regions), faculae (bright filaments), and other features.
One technique astronomers have used to try to model the stellar activity component of the signal is principal components analysis (PCA), applied to an ensemble of spectra collected over time.
But PCA neglects the continuous function nature of a spectrum at a given time, and it takes no account of the continuous evolution of spectra over time.
We address model building from a function approximation perspective, via the asymmetric Hilbert-Schmidt decomposition (HSD).
This provides a PCA-like low-rank decomposition of a dynamic spectrum, but it explicitly accounts for time dependence, both estimating temporal basis functions, and accounting for irregular temporal sampling when estimating spectral basis functions.
We show that HSD improves the fidelity of a fixed-order model of the dynamic spectrum of a Sun-like star with a starspot, with mean squared error for spectral basis functions (vs.\ optimal basis functions) an order of magnitude or more times smaller than that achieved via PCA when spectra are randomly sampled sparsely and irregularly in time.

\end{abstract}


%===============================================================================
%===============================================================================

\section{Introduction}
\label{sec-intro}

Spectroscopy---measurement of the intensity of light as a function of wavelength or frequency---provides fundamental insights into diverse physical, chemical, and biological phenomena.
Its impact on a field can be transformative.
In astronomy, for example, the development of stellar spectroscopy in the nineteenth century is credited for the birth of astrophysics, as it identified the compositions of stars, a key foundation for the theory of stellar structure.
In the twenty-first century, \emph{time-resolved spectroscopy} has emerged as a powerful tool for probing the dynamics of many systems---e.g., stellar variability, the molecular dynamics of complex molecules, and fast chemical processes in biological systems.
The work reported here is motivated by searches for Earth-like planets around Sun-like stars using the radial velocity (RV) method, which detects a  planet indirectly by measuring the minute reflex motions it induces on its host star via the time-dependent Doppler shift of the star's spectrum.
Intrinsic variability of the host star spectrum due to stellar activity (e.g., starspots) can mask or mimic the minute time-dependent Doppler shift signals that small planets produce in such data, and is a main obstacle to discovering Earth-like exoplanets, a major goal of extreme precision radial velocity (EPRV) surveys \citep{F+16-EPRV,BDH26-PreciseRVs}.
Confident detection requires tailored but flexible modeling of the spectral signature of stellar activity to disentangle it from a genuine exoplanet signal; this remains an open problem.

Measurements of a time-resolved spectrum (also called a \emph{dynamic spectrum}) comprise a type of spatiotemporal data, with photon wavelength, frequency, or energy serving as the (univariate) spatial dimension; we use wavelength here (the conventional choice for stellar spectroscopy).
A key feature of most time-resolved or dynamic spectral data is \emph{strong asymmetry between the spectral and temporal sampling}.
The spectral dimension is typically finely sampled on a regular grid of wavelengths (or reciprocal or logarithmic wavelengths), with the grid comprising hundreds to hundreds of thousands of wavelengths, depending on the application.
In contrast, the temporal dimension is typically less fully sampled, with perhaps tens to $\sim 10^3$ samples.
In some applications---including the EPRV application treated here---the temporal sampling is very irregular.
And when multiple, related objects or systems are studied, the temporal sampling may differ across the objects.
We describe an approach for modeling dynamic spectra devised specifically to accomodate and adapt to this spectral/temporal sampling asymmetry, in particular, to account for irregular temporal sampling.

A time-resolved (dynamic) spectrum can be described with a bivariate function of wavelength and time, $\aspec(\wob,\tob)$, representing the intensity of light measured by a spectrograph (e.g., the time-dependent photon flux density, an intensity function for a Poisson process governing photon detections), or a (relative) measurement of absorption or emission vs.\ wavelength, $\wob$, and time, $\tob$.
The task is to model this function in a way that enables separation of an exoplanet Doppler shift component from spectrotemporal behavior due to stellar variability.
If dynamic spectra were available from many similar stars without planets, from a statistical perspective, \emph{functional data analysis} (FDA)---the body of techniques for modeling data from measurements of an population of related functions---provides an ideal framework for characterizing stellar activity across a stellar population.
But planet searches sensitive to planets more massive than Earth indicate that most stars are likely to host planets.
Ab initio characterization of stellar activity via FDA (or other similar approaches) would require joint modeling of spectral variability, and detection and characterization of planets unique to each star in the studied population---a daunting task.

If one focuses on planet searches targeting Sun-like stars (i.e., G dwarfs), the Sun provides a workaround.
Its spectrum has been observed carefully for decades.
We have detailed knowledge of the orbits of the planets in the solar system, so planet-induced signals in the solar spectrum can be removed with high accuracy and precision, providing access to high-precision measurements of G dwarf stellar activity.
The long duration of measurements provides access to diverse states of solar variability, which can serve as a stand-in for observing a population of similar stars for shorter durations.
Finally, the Sun's proximity enables observation of the solar spectrum at extremely high signal-to-noise; the data are effectively noiseless.
For these reasons, a significant body of work on stellar activity modeling for EPRV planet searching has focused on modeling solar activity in a way that could translate to modeling dynamic spectral data from other G dwarfs (and perhaps more broadly).

A common theme in the literature on stellar activity modeling is \emph{dimension reduction}---finding a small set of features derived from stellar spectra that capture most of the impact of stellar variability on the evolving spectrum \citep{HF23-ExoRV-ARSA}.
Early work attempted to use spectral features empirically found to characterize solar variability over decades of observational studies by solar astronomers.
Such features are helpful for modeling variability at the level needed to enable detection of massive planets, but are not adequate in the EPRV regime targeting detection of Earth-mass planets.
Recent work uses statistical and machine learning algorithms to implement dimension reduction in a more automated, data-driven way.

An influential body of work applies principal components analysis (PCA) to data-based simulated solar spectra to study and model solar variability (see the recent reviews by \citet{HF23-ExoRV-ARSA,BDH26-PreciseRVs}).
The spectrum at a given epoch (observing time) is treated as a vector of multivariate data, and PCA is applied to the collection of spectra observed over time to identify a small set of principal component (PC) spectra whose time-dependent linear superposition captures most of the spectral variability.
\citet{D+17-SOAP+PCA} used this approach to quantify the complexity of solar variability.
Building on their work, several other teams have similarly used PCA, applied directly to spectra or to related observations (e.g., spectrum autocorrelation functions, or estimated spectrum derivatives) to model variability in Sun-like stars (in some cases attempting to extend the approach other stellar types); see, e.g., \citet{C+21-RVviaACF+PCA,J+22-ExoDtxn-PCA+GP,CDP22-PCAShells}.

From a statistical perspective, this work raises conceptual issues whose resolution can impact calculations.
PCA is a method from multivariate statistics; its rationale and optimality follow from the vector nature of multivariate data, and from considering the collection of measurements to be independent and identically distributed (IID) draws from a finite-dimensional multivariate distribution.
Considering stellar dynamic spectra instead as described by a spectrotemporal stochastic process, whose statistical properties derive from the dynamic physics of stellar photospheres, challenges these assumptions.
Spectral measurements at a particular epoch represent measurements of a continuous function, not of a finite-dimension vector.
Further, spectra measured across time should be statistically dependent, in a way that is tied to the temporal sampling.
In the case of the simulated solar spectra used for multiple studies, which are based on effectively noiseless measurements of solar spectra, the dynamic spectrum is in fact a smooth two-dimensional deterministic function; stochasticity across measurements of dynamic spectra of similar stars reflects star-to-star variability in physical parameters and initial conditions (e.g., the sizes and locations of active regions on each star).
Despite the conflict between the nature of stellar spectral variability and the assumptions behind PCA, PCA-based modeling of stellar activity is empirically doing something useful.
Our goal is to establish a sound rationale for PCA-like empirical basis function modeling of dynamic spectra, leading to a framework that improves on PCA, and points to generalizations that could be even more useful for EPRV data analysis.

The work reported here examines stellar activity model building from the perspective of \emph{function approximation theory} rather than multivariate statistics, to build a foundation for stochastic process models that could model activity from an FDA-based statistical perspective.
The key tool we rely on is the \emph{asymmetric Hilbert-Schmidt decomposition} (HSD), an approximation theory approach for building separable expansions for bivariate functions using univariate empirical basis functions in each variable (here, wavelength and time).
Historically, the development of the HSD provided the foundation for the singular value decomposition (SVD), the linear algebra tool often used to compute PCA decompositions \citep{S93-HistorySVD,S11-FredholmHilbertSchmidt}.
Computing the HSD in practice thus uses similar linear algebra to that used for PCA.
But there are important differences in the calculations, reflecting that HSD seeks a functional rather than a finite-dimensional multivariate decomposition.
One manifestation of this is that the time separations between spectral observations play an explicit role in the HSD; they are ignored by PCA.

The next section presents the EPRV planet detection problem, describing the exoplanet and stellar variability signals; it also describes the stellar activity simulation tool we use in this study, the Spot Oscillation And Planet (SOAP) code \citep{BBS12-SOAP,DBS14-SOAP2,C+25-SOAPv4}.
% As mathematical background, \S~\ref{sec-decomp} describes three decompositions---PCA, the Karhunen-Loeve expansion (KLE) for stochastic processes (the foundation for functional principal components analysis, FPCA), and the HSD.
% This section aims to clarify the conceptual basis for each of these related decompositions.
\S~\ref{sec-decomp} describes recent approaches for accounting for stellar variability, highlighting the role of PCA in recent work, and using it to motivate modeling a dynamic spectrum with a low-rank separable expansion computed via HSD.
\S~\ref{sec-comparison} compares how accurately PCA and HSD can describe stellar activity due to a sunspot, demonstrating that HSD offers significant improvements over PCA, particularly when temporal sampling is irregular.
Finally, \S~\ref{sec-summary} summarizes the main results and briefly describes how the HSD and related separable functional expansions could be used as a foundation for stochastic process modeling of EPRV data, including from an FDA perspective.

\section{EPRV planet searching and SOAP simulations}
\label{sec-EPRV}

%-------------------------------------------------------------------------------
\subsection{Doppler shift}

The RV method detects and characterizes planets by measuring the line-of-sight reflex motion of a star that may host exoplanets, via the time-dependent Doppler shifts the motion induces in the star's face-averaged spectrum, measured via the locations of hundreds to thousands of absorption lines in the spectrum.
Modern spectrographs count photons, and the incident spectrum at the top of Earth's atmosphere can be described by the photon number flux density (also called the specific photon number flux, i.e., specific in wavelength), $\pflux_\wob(\wob,\tob)$, specifying the expected number of photons arriving per unit area (normal to the star's direction), unit time, and unit wavelength (the $\wob$ subscript is a convention distinguishing flux per unit wavelength from flux per unit photon frequency, $\nu$).
The expected number of photons detected in a spectrograph pixel depends on the duration of the observations, the area of the pixel (as seen through the telescope), and the spectral bandwidth covered by the pixel.
It also depends on the throughput of the telescope and spectrograph optics, and the quantum efficiency of the detector pixels.
It further depends on the transparency of the atmosphere along the line of sight, which varies strongly with wavelength (due to absorption by various atomic and molecular species), and which varies significantly with time.
Spectrophotometric measurements, which aim to accurately recover both the shape and amplitude of $\Phi_\wob(\wob,\tob)$, are challenging, requiring constant calibration of the instrument and telescope, and constant monitoring of atmospheric conditions (e.g., via near-simultaneous measurement of calibration sources).
Most astronomical spectroscopy is not spectrophotometric; the measurements accurately recover the shape of the flux density, locally in wavelength and time, but cannot recover the amplitude with high accuracy consistently across a measured spectrum, or between spectra measured at well-separated times.
EPRV measurements are not spectrophotometric.

Reflecting this, reported spectra aim to accurately capture only the locations and shapes of line features.
The reported spectra are found by adjusting measured spectra using an estimate of the local, smooth continuum spectrum (i.e., a smoothed version of the spectrum with lines masked out).
Different observing teams perform the adjustment in different ways (e.g., using different methods to estimate the continuum and factor it out from the estimated spectrum).
We refer to the reported spectrum as an \emph{adjusted spectrum} (a more common term in the EPRV literature is \emph{normalized spectrum}, but such spectra are not ``normalized'' in the statistical sense of having unit area, so we eschew that term here).
Henceforth, ``spectrum'' refers to the adjusted spectrum unless specified otherwise.

\begin{figure}[t]
\centerline{\includegraphics[width=1.1\textwidth]{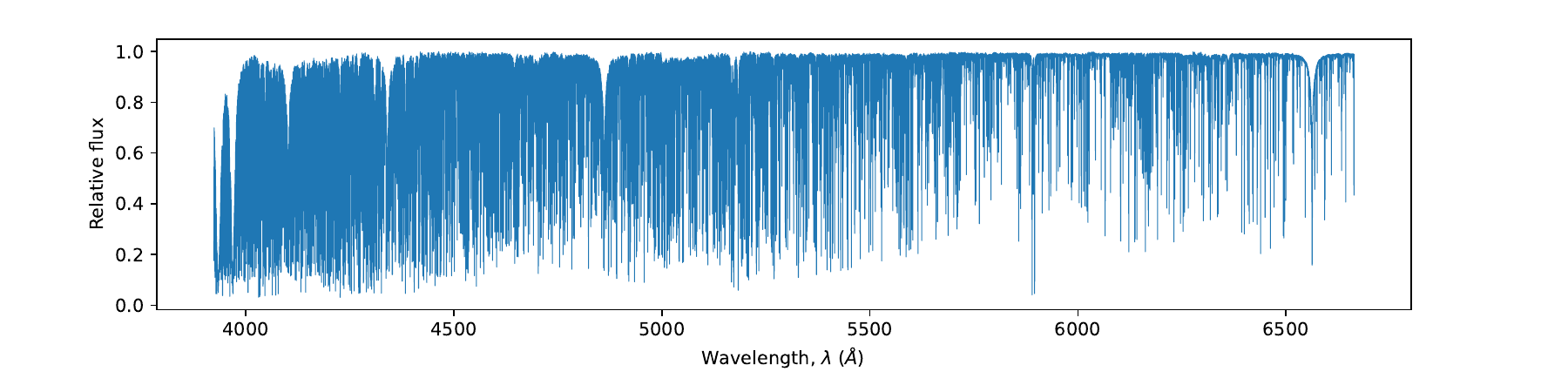}}
\caption{
Face-integrated adjusted spectrum of the quiet Sun, as measured by the NSO FTS.
The curve shows the relative amplitude of the solar spectrum, with respect to the locally estimated continuum, as a function of wavelength in angstroms.
Thousands of absorption lines are visible.
\label{fig:SOAPSpec}}
\end{figure}

\cfig{fig:SOAPSpec} shows an (adjusted) spectrum from observations of the quiet Sun (i.e., with no significant solar activity).
The smoothed flux density of the Sun is very roughly that of a black body with a temperature of 5800~K, varying strongly with wavelength with a peak near 5000~\AA.
The adjusted spectrum factors out that smooth dependence, so its upper envelope is nearly flat, set to unity by convention.
Thousands of absorption lines are evident in the spectrum, varying greatly in width, depth, and shape.
(\cfig{fig:SOAPSpecZoom}, discussed below, provides a sequence of zoomed-in views of this spectrum.)

Focusing for the moment solely on wavelength dependence, let $\aspecem(\wem)$ denote the spectrum in the rest frame of the star, at rest-frame wavelength $\wem$, and $\aspec(\wob)$ denote the spectrum in the observer rest frame, as a function of observer-frame wavelength $\wob$.%
\footnote{Here ``observer rest frame'' or ``observer frame'' refers to an inertial (non-accelerating) frame adopted by observers.
By convention this is the reference frame of the solar system barycenter (the center of mass of the solar system).
Earth-bound observers adjust their raw measurements to correspond to barycentric measurements, which removes impacts on the spectrum from the rotation of the Earth, the revolution of the Earth in its orbit, and the reflex motion from the revolution of the other planets (and large minor planets) in their orbits.}
A photon emitted with wavelength $\wem$ in the frame of reference of a moving star will be observed to have wavelength
\be
\wob = \frac{\wem}{\dop},
\label{w-w-dop}
\ee
where $\dop$ is the \emph{Doppler factor}, a function of both the line-of-sight (radial) velocity of the star with respect to the observation reference frame, $\vlos$, and the total velocity, $v$:
\be
\dop = \frac{\sqrt{1 - (v/c)^2}}{1 + \vlos/c},
\ee
with $c$ denoting the speed of light.
The denominator  accounts for the bunching or spreading of wavefronts as the source moves toward or away from the observer (negative or positive $\vlos$, resp.); the numerator accounts for relativistic time dilation.
It is common to describe the wavelength transformation in terms of the \emph{redshift}, $\zz$, defined as the fractional change in wavelength,
\be
\zz \equiv \frac{\wob - \wem}{\wem} = \frac{1}{\dop} - 1,
\label{z-def2}
\ee
so that $1/\dop = 1 + \zz$ and $\wob = (1+z)\wem$.
Thus the observed and rest-frame spectra are related according to
\be
\aspec(\wob) = \aspecem\left(\wob/(1+z)\right).
\label{aspec-z}
\ee
In the nonrelativistic regime relevant to most exoplanet work, the dependence of $\dop$ on total velocity can be ignored (it is of second order), and $1+z \approx 1+\vlos/c$.
Thus to leading order, $\wob/(1+z) \approx \wob(1-z)$, and
\be
\aspec(\wob) \approx \aspecem(\wob - z\wob).
\label{aspec-z-lin}
\ee
The Doppler shift is more accurately a dilation rather than a shift (translation), but \ceqn{aspec-z-lin} reveals that for small redshifts (line-of-sight velocities small with respect to $c$), the dilation amounts to a local shift (albeit proportional to wavelength).%
\footnote{The kinematical considerations that underly the Doppler shift in wavelength also affect the amplitude of the spectral flux density, $\pflux_\wob$.
For the small values of redshift relevant for exoplanet searches, the Doppler effect on the amplitude of the flux density is much too small to be measurable with respect to instrument calibration uncertainties, even with spectrophotometric observations.
In any case, they are not measurable using adjusted spectra, so we ignore them here.}

In exoplanet settings, the relevant line-of-sight velocities are $\sim 10~\text{cm s}^{-1}$ (for Earth-like planets) to $\sim 1~\text{km s}^{-1}$ (for Jupiter-sized planets close to their hosts, the prevalent ``hot Jupiters'').
With $c\approx 3\times10^5~\text{km s}^{-1}$, this corresponds to redshifts $z \sim 3\times10^{-10}$ to $3\times10^{-6}$, corresponding to spectral line shifts of $\sim 10^{-6}$ to $10^{-2}$~\AA\ for spectral lines near 5000~\AA\ (the optical spectrum spans $\approx 3800$ to 7500~\AA).
For small planets this corresponds to shifts less than 1/1000th of the typical width of a spectral pixel in EPRV observations.
Measuring such minuscule shifts is achievable only by pooling information across hundreds to thousands of spectral lines measured at high signal-to-noise (i.e., from relatively nearby stars).

%-------------------------------------------------------------------------------
\subsection{The radial velocity (RV) method}

The RV method uses spectra measured at various epochs (observing times), $\tob_k$ (for $k=1$ to $K$), to estimate the host star redshift as a function of time, $z_k \equiv z(t_k)$.
Planet detection and characterization proceed by subsequent analysis of the time series of estimated redshifts, e.g., by looking for periodic behavior of redshift vs.~time, with $z(t_k)$ having a shape consistent with the predictions for Keplerian orbits (or, in some systems with multiple, interacting planets, with predictions from numerical $N$-body calculations).

Observed (adjusted) spectra may be modeled as measurements, at each epoch $t_k$, of the spectrum sampled at wavelengths $\wob_{ik}$ ($i=1$ to $N_k$), with additive zero-mean Gaussian noise with known standard deviations:
\begin{align}
Y_{ik} 
  &= \aspec(\wob_{ik}) + \serr_{ik}; \qquad \serr_{ik} \sim N(0,\sigma_{ik})\nonumber\\
  &= \aspecem(\wob_{ik} - z_k\wob_{ik}) + \serr_{ik}.
\label{Y-def}
\end{align}
% (Measurements at different epochs are typically available at epoch-dependent wavelengths, $\wob_{ik}$; we suppress the $k$ dependence here for simplicity.)
If the rest-frame adjusted spectrum, $\aspecem(\wem)$, were known, we could estimate $z_k$ independently at each epoch via straightforward nonlinear regression based on \ceqn{Y-def}, considering $z_k$ to be a single unknown parameter.
In fact, since $z\ll 1$, we can linearize \ceqn{Y-def}, writing (for each epoch)
\be
Y_{ik} = \aspecem(\wob_{ik}) - z_k\wob_{ik} \daspecem(\wob_{ik}) + \serr_{ik},
\label{Y-F0-lin}
\ee
with $\daspecem$ denoting the derivative of the rest-frame adjusted spectrum.
With $\aspecem(\wem)$ and its derivative known, estimation of the redshift at a particular epoch could in principle be done via linear regression.

In reality, $\aspecem(\cdot)$ is \emph{not} known.
% Consider for the moment a star with no stellar activity.
Schematically, we may write
\be
Y_{ik} = \aspecem(\wob_{ik};\spar) - z_k\wob_{ik} \daspecem(\wob_{ik};\spar) + \serr_{ik},
\label{Y-F0-lin-spar}
\ee
where $\spar$ denotes a vector of parameters describing the rest-frame adjusted spectrum (the derivative, $\daspecem$, is now a partial derivative with respect to wavelength).
The $\spar$ vector in such a formulation of the problem would be high-dimensional (e.g., it would have to desribe the locations and shapes of thousands of lines).
Estimating the redshifts at all epochs would need to be done jointly (since each epoch contributes information about $\spar$), and would require solving a high-dimensional regression problem, with the $\spar$ parameters being nuisance parameters.

Instead of formally following such a regression approach, astronomers mainly rely on clever ad hoc methods for estimating $z_k$ at a given epoch.
% We briefly describe these approaches in \S~\ref{sec-comparison}.
The most-used methods adopt a template of some sort, and estimate $z_k$ by comparing redshifted versions of the template to the observed spectrum \citep{F+16-EPRV,LS26-ExoDtxnTechRV}.
The \emph{cross-correlation function (CCF) method} uses a mask constructed from known line positions as the template; the mask may be binary (with unit positive spikes at nominal line positions) but more typically is weighted (to enable some lines to impact the estimate more than others).
The mask is stepped in redshift and the correlation between the mask and the observed spectrum is computed; the minimum of the CCF gives the estimated redshift (the minimum rather than the maximum is relevant because the mask has positive spikes, and the spectrum has absorption lines).
The \emph{template matching method} computes a detailed, high signal-to-noise spectral template for the host star spectrum by combining multiple spectra.
Redshift is estimated via a statistical comparison of redshifted versions of the template to the observed spectrum at a specific epoch, e.g., via least squares or Bayesian fitting \citep{S+22-SemiBayesRV}.
This is close in spirit to the regression approach described above, though it does not jointly fit the rest-frame spectral shape with redshifts, and present implementations account for spectral shape uncertainty in the redshift estimates in a limited way (e.g., by adding a pointwise template variance term to the weights for weighted least squares fitting).

%-------------------------------------------------------------------------------
\subsection{Impact of stellar activity}

A further challenge---and our focus here---is that stellar activity makes the rest-frame spectrum \emph{evolve} in subtle and complex ways.
That is, the rest-frame spectrum is really a function of both (rest-frame) wavelength and time, $\aspecem(\wem,\tem)$, and the data at epoch $t_k$ depend on time, not only via $z(t_k)$, but also due to spectral evolution.%
\footnote{The rates of time in the observer and rest frames are also impacted by the kinematics underlying the Doppler shift: $\dif t = \dif t_0/\dop$.
For the very small redshift regime relevant to exoplanet searches, the Doppler dilation of time is not significant, and henceforth we consider spectra only as functions of observer time.}
The regression formulation outlined above thus needs to be generalized to include time dependence of the rest-frame spectrum:
\be
Y_{ik} = \aspecem(\wob_{ik}, t_k;\spar) - z_k\wob_{ik} \daspecem(\wob_{ik},t_k;\spar) + \serr_{ik},
\label{Y-F0-lin-spar-time}
\ee
where now the rest-frame spectrum parameters, $\spar$, specify the dynamic rest-frame spectrum, a bivariate function of $(\wob,t)$.
The main goal of this work is to establish a conceptual and mathematical basis for building such models, from bivariate function approximation theory.

\begin{figure}[t]
\centerline{\includegraphics[width=.9\textwidth]{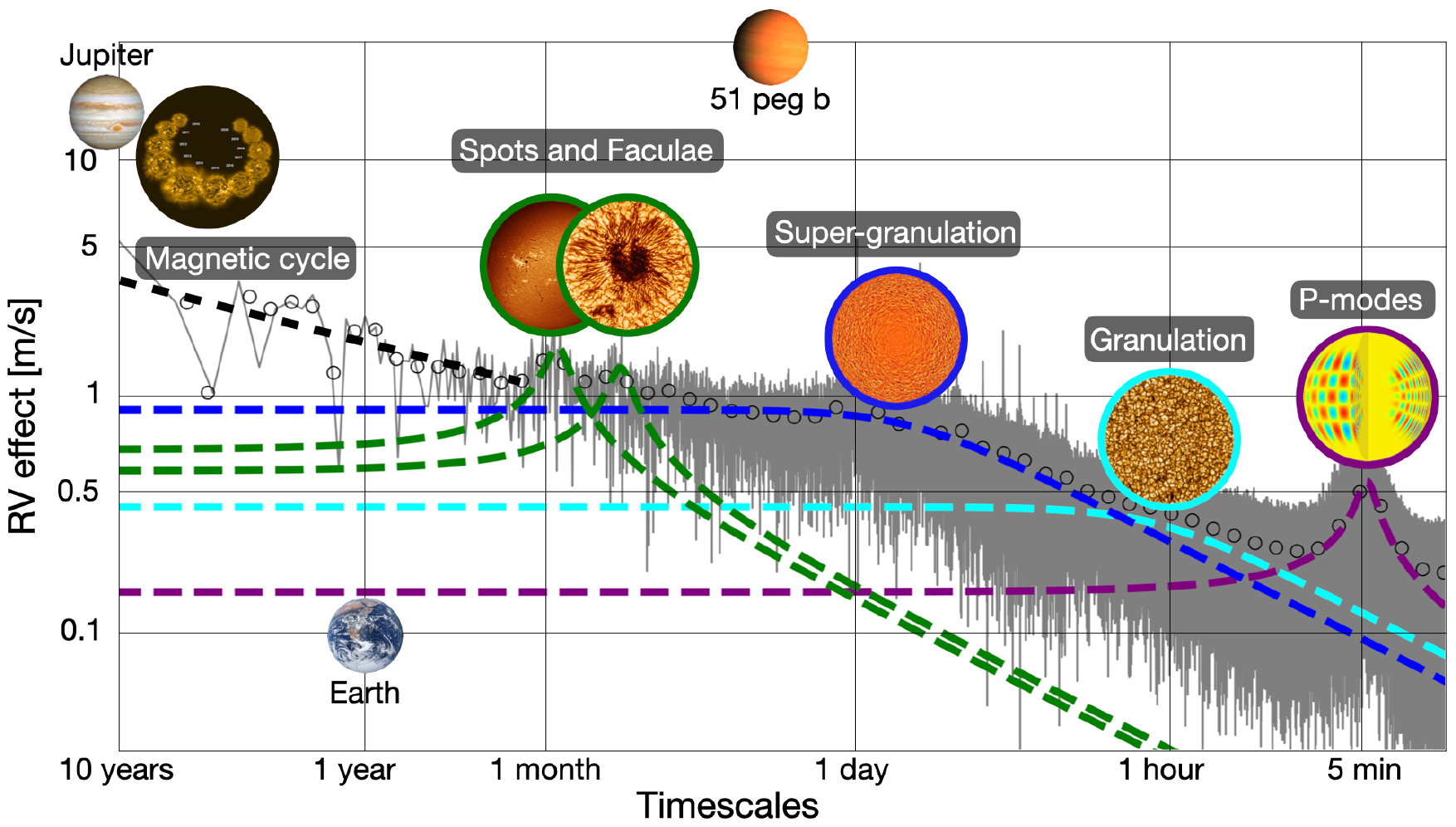}}
\caption{
Radial velocity amplitude of solar variability vs.\ time scale. 
Gray curve shows the measured Fourier power spectral density; colored dashed curves show fitted contributions for different sources of variability.
Annotations label the dashed curves with depictions of phenomena causing variability, and indicate the RV amplitudes and orbital periods of stellar reflex motion in response to the orbits of the planets Earth and Jupiter, and exoplanet 51~Peg~b (a hot Jupiter).
From \citet{BDH26-PreciseRVs}.
\label{fig:StellarVar}}
\end{figure}

\cfig{fig:StellarVar} shows how various solar surface and subsurface phenomena comprising stellar activity impact the face-averaged radial velocity of the Sun over a wide range of time scales (the figure is from the EPRV review article by \citet{BDH26-PreciseRVs}).
The gray curve (fluctuating around the small black circles) shows the Fourier power spectral density for face-averaged solar RV measurements over time (the RV measurements are with respect to the solar system barycenter, i.e., with reflex motion due to the planets removed).
The colored dashed curves show fitted contributions for different sources of variability, depicted with graphical annotations.
Granulation and supergranulation refer to large-scale convective motions happening on and just below the photosphere, involving regions with length scales from hundreds to tens of thousands of kilometers, with local motions of up to a few \kmps.
The net variability, after averaging over the face of the Sun, can be $\sim 1$~\mps.
Seismic activity in the outer layers of the Sun produce p-mode (pressure-mode) oscillations with time scales of several minutes (these are sound waves in the solar plasma).
By averaging observations over times longer than this, the impact of p-mode variability can be largely mitigated.
Sunspots (localized regions of low effective temperature) and faculae and plages (lanes of high temperature plasma in the photosphere and chromosphere, resp.) persist over time scales of weeks to months, slowly evolving; they impact the time dependence of face-averaged RV measurements largely due to solar rotation, which has a period of $\approx 25$ days at the solar equator and $\approx 36$ days at the solar poles.
Spots, faculae, and plages are associated with the Sun's magnetic activity, which varies over time scales of years to centuries and probably on much longer time scales (there is quasiperiodic magnetic variability, with the Sun's magnetic field flipping polarity every $\approx 22$ years; it is synchronized with quasiperiodic variability in the number of sunspots with a period of $\approx 11$ years).
Stochastic variability, as from granulation, can mask the RV signal of a small planet.
Quasiperiodic variability, as from sunspots and faculae, is especially problematic as it can mimic the RV signal of a small planet if the span of observations does not extend over multiple rotation periods (so that nonperiodic evolution of the activity becomes evident).

\begin{figure}[t]
\centerline{\includegraphics[width=.7\textwidth]{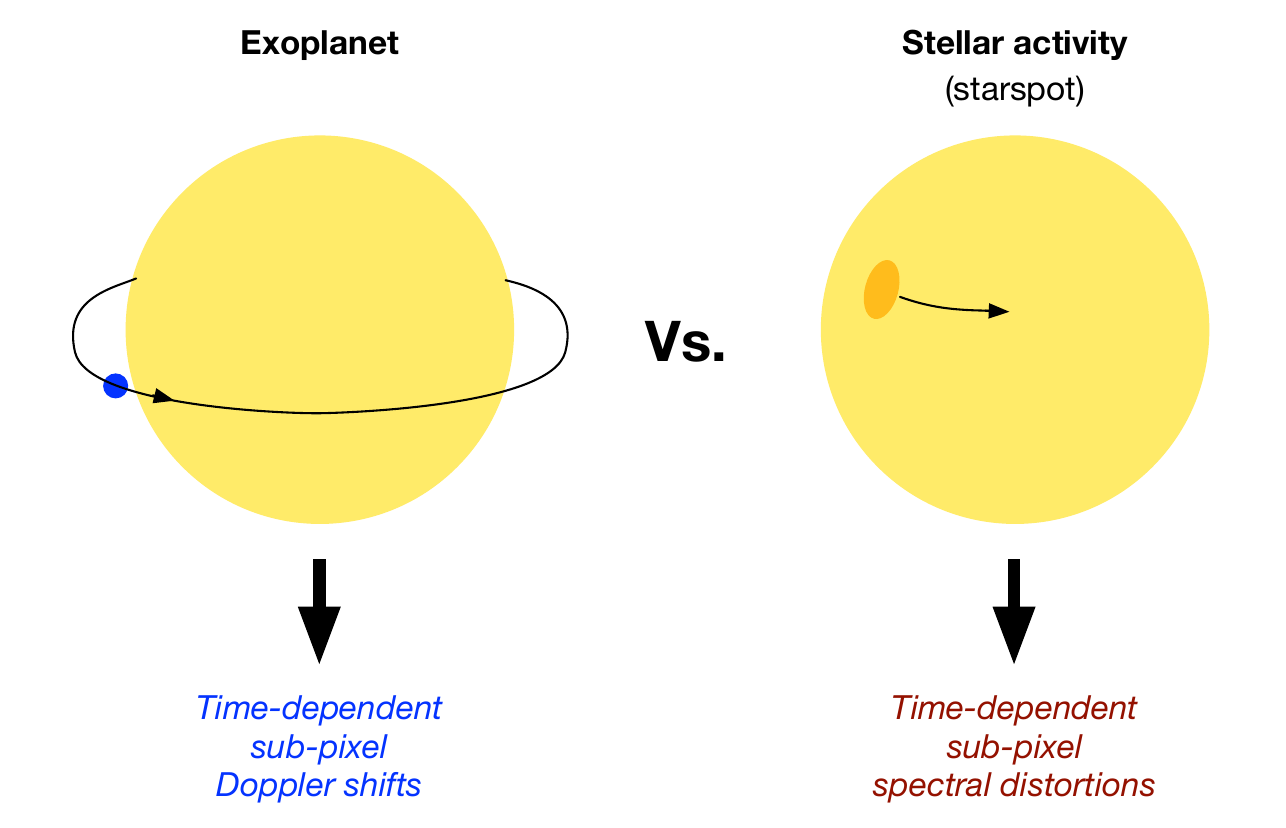}}
\caption{
Illustrations depicting the impact of an exoplanet and a starspot on the dynamic spectrum of a star.
\label{fig:ExoVsSpot}}
\end{figure}

We focus here on modeling variability due to stellar spots.
\cfig{fig:ExoVsSpot} provides a graphical depiction of how spots are problematic.
The left panel depicts a planet orbiting a star.
Only rarely can we directly image planets orbiting a star outside the solar system (for nearby systems, with large planets in distant orbits, and favorable orientations).
The RV method detects planets indirectly via their impact on the line-of-sight velocity of the host star, which is not spatially resolved, so the RV measurement is a face-averaged measurement.
As the planet (which is not directly observable) approaches the observer, the host star recedes, exhibiting periodic Keplerian reflex motion (the ratio of the semimajor axes of the orbits of the planet and star about their center of mass is the inverse of the ratio of their masses).
This causes Doppler shifts in the spectral lines observed from the star.
The right panel depicts a star with no planet, but with a large stellar spot.
As the dark spot comes over the stellar limb, approaching the observer, it effectively subtracts off blue-shifted light from the approaching limb from the face-averaged spectrum used for RV measurement.
The effect evolves, eventually changing sign as the spot rotates across the face of the star (effectively subtracting red-shifted light).
In this way, spots produce nearly periodic changes in the spectrum of a star that corrupt RV measurements, potentially mimicking a planetary signal.

\begin{figure}[t]
\centerline{\includegraphics[width=\textwidth]{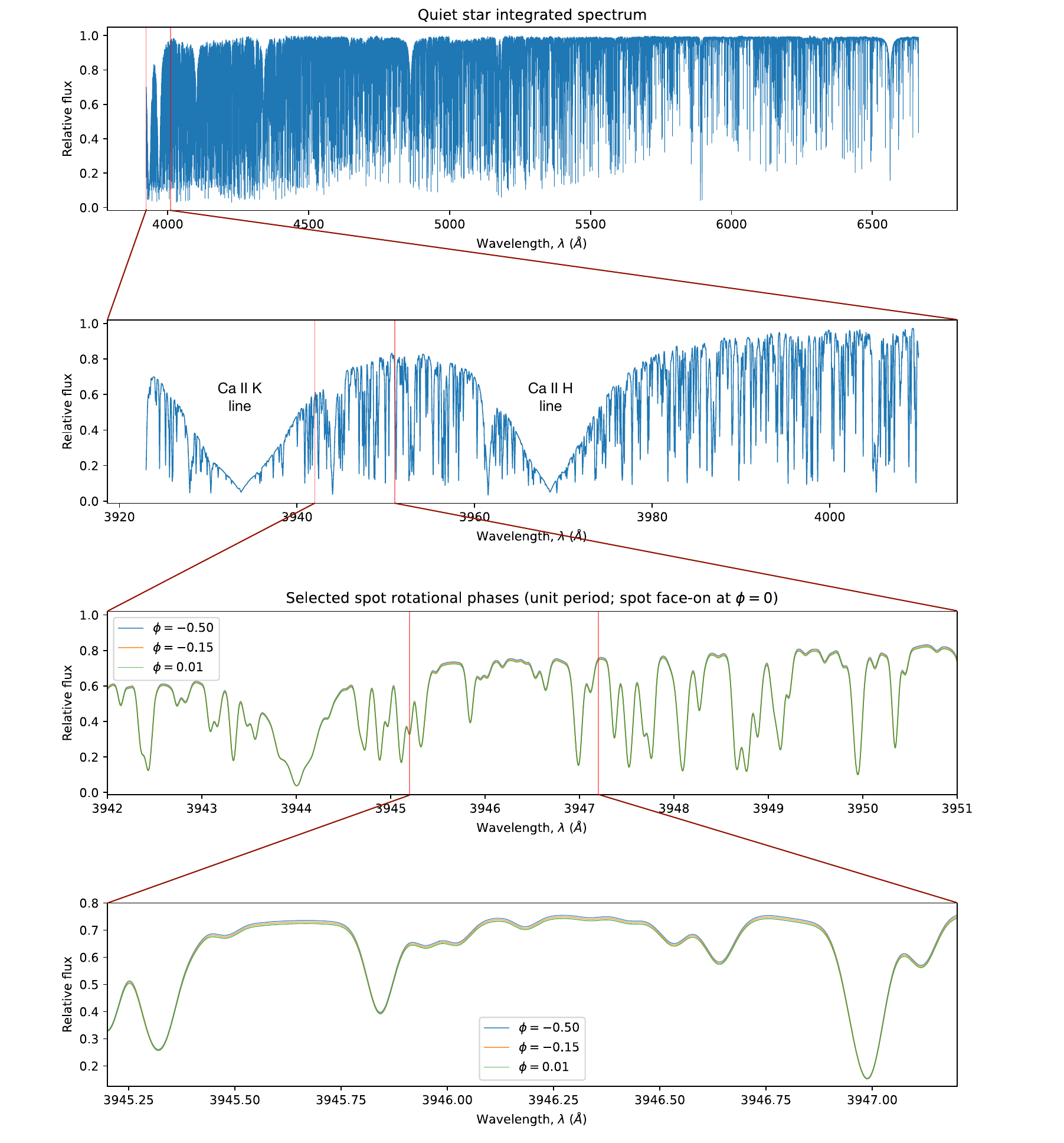}}
\caption{
Time-evolving spectrum of a Sun-like star with a single equatorial starspot occupying 1\% of the stellar surface, generated using SOAPv4.
Top two plots show the quite-star spectrum (i.e., with the spot hidden), over the full optical range (top panel), and zoomed in to the vicinity of the Ca~II H and K lines (2nd from top).
The bottom plots each show three spectra, at further zoom levels, with the spot at three rotational phases, $\phi$, as indicated.
\label{fig:SOAPSpecZoom}}
\end{figure}

\cfig{fig:SOAPSpecZoom} shows the effect of a spot on a stellar spectrum in detail.
It shows spectra from simulations of a single equatorial starspot occupying 1\% of the stellar surface of a Sun-like star, at various phases of stellar rotation; it was produced using the SOAPv4 code (described below).
The top panel shows the quiet-star spectrum (with the spot on the back hemisphere, thus out of sight), for reference.
The next panel zooms in to the vicinity of the broad Ca~II (single-ionized calcium) H and K lines (corresponding to different electron transitions).
The bottom two panels each show three spectra at two further zoom levels. 
The three spectra have the spot at three rotational phases, $\phi$ (fractional rotation in $[-0.5, 0.5]$)---spot hidden at phase  $\phi=-0.5$ (as in the top two panels), spot face-on at $\phi=0$, and an intermediate case with $\phi=-0.15$.
The three spectra only become distinguishable at the highest zoom level:
the bottom panel reveals a barely discernable reduction in the
spectrum amplitude with the spot present, due to the spot's reduced temperature (compared to that of the quiet photosphere).

\begin{figure}[t]
\centerline{\includegraphics[width=0.9\textwidth]{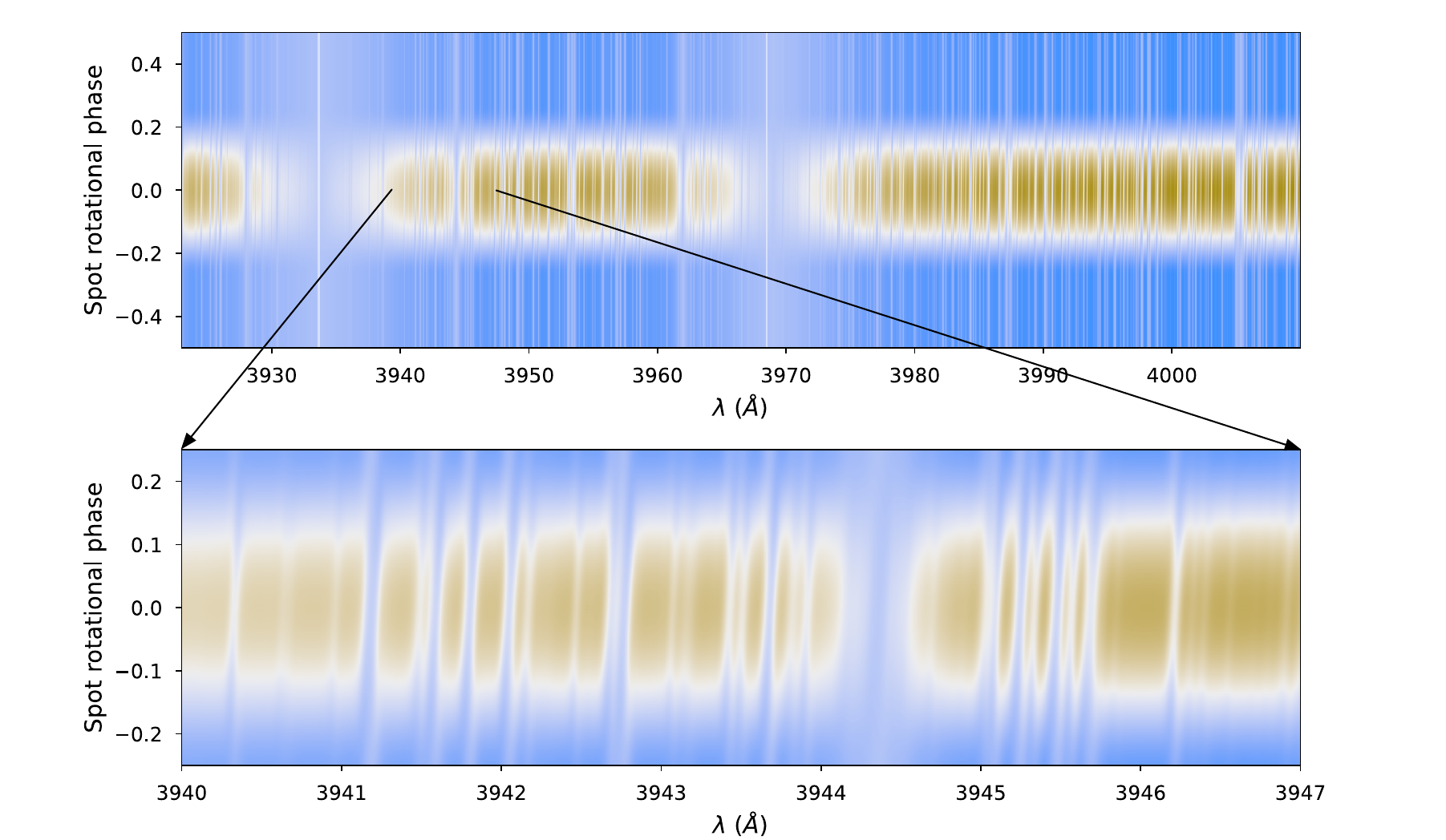}}
\caption{
SOAPv4 spectra as in \cfig{fig:SOAPSpecZoom}, but here for 100 phases, displayed as difference images. 
Time (phase) is vertical, and wavelength horizontal.
The color map uses brown hues (``ground'') for negative values, white hues for values near 0, and blue hues for positive values (``sky'').
The bottom panel zooms in on the indicated region in the top panel.
 \label{fig:SOAPSpecDiffImage}}
\end{figure}

\cfig{fig:SOAPSpecDiffImage} shows the same evolving spectrum, but now computed at 100 evenly-space phases in the interval $[-0.25, 0.25]$, corresponding to the spot just becoming visible, then face-on, and then rotating out of sight.
To visualize the small changes in the spectrum, \cfig{fig:SOAPSpecDiffImage} displays the spectra as a difference image, showing the difference between the instantaneous spectrum and the time-averaged spectrum, for the wavelength regions covered in the two middle panels of \cfig{fig:SOAPSpecZoom}, as a false-color image (time increases from bottom to top).

\unskip

The difference image reveals that different regions of the spectrum are affected by the spot in different ways, even in opposite senses.
Particularly notable are the slanted streaks evident at high zoom level. 
These correspond to small, systematic shifts in line locations with time---spectral distortions that resemble a Doppler shift, with the potential to corrupt stellar RV measurements in exoplanet searches.
These distortions largely come from the spot slightly masking blue-shifted stellar emission as the spot emerges over the advancing limb (where the stellar surface is moving toward the observer), and later masking red-shifted emission as it moves to the receding hemisphere.

%-------------------------------------------------------------------------------
\subsection{SOAPv4}

We computed the spectra displayed in the figures above using SOAPv4, a recent open-source update of the Spot Oscillation And Planet (SOAP) code widely used to simulate realistic stellar spectra, including stellar activity signatures from spots or faculae \citep{BBS12-SOAP,DBS14-SOAP2,C+25-SOAPv4}.
We also used SOAPv4 for more extensive simulations of star-plus-spot dynamic spectra used for the calculations described below.
The SOAPv4 algorithm pixelates a hemisphere representing the visible face of a star, populates the pixels with stellar spectra as prescribed by the user, and computes the face-averaged spectrum (and related summary statistics), accounting for pixel-dependent geometric projection, Doppler shift, and limb darkening (the impact of viewing angle on radiative transfer in the outer layers of the photosphere).
The user specifies the orientation and rotational period of the star, the quiet stellar spectrum, the location and size of one or more circular active regions (spots or faculae), and spectra for the active regions.
Optionally, the effect of a planet can be included by specifying the planet's orbital parameters, mass, and radius.
SOAP then accounts for the reflex motion Doppler shift on the spectrum, and can also compute the effects of a transit when the orbital geometry leads to the planet crossing the observed face of the star.

% TODO:  Improve description of early SOAP use of CCF.
% TODO:  Give values for resultion and sampling of FTS spectra.
% TODO:  Add FTS data refs.

Early versions of SOAP used only spectral summaries in each pixel (essentially, measurements of average line shapes).
Beginning with SOAP~2.0 \citep{DBS14-SOAP2}, users could choose to use archival observed solar spectra for the quiet star pixels and for spot pixels.
SOAPv4 provides access to the quiet Sun spectra measured by the Fourier transform spectrograph (FTS) at the Institut für Astrophysik und Geophysik (IAG) in Göttingen, and to both quiet Sun and spot spectra measured by the FTS at the National Solar Observatory (NSO) in Boulder, Colorado.
Both the IAG and NSO spectra are measured at very high resolution, finely sampled (at $\approx$~500,000 wavelengths for the NSO spectra) and with very high signal-to-noise ($\sim 1000$), thus providing the capability to model spectra with starspots for Sun-like stars with high fidelity.
SOAPv4 also provides access to pre-computed stellar spectra produced by leading stellar atmosphere codes.
These enable simulation of dynamic spectra for stars other than Sun-like G dwarfs.

For the calculations reported here, we used $3.6\times 10^5$ pixels, a stellar orientation corresponding to the rotation axis being in the plane of the sky, and the NSO FTS spectra.
We focus on activity due to spots (deferring consideration of faculae for future work), considering spots of various size and at various solar latitudes, as specified below.
These choices are very similar to those used in previous work using PCA and data simulated with SOAP~2.0 (e.g., \citet{D+17-SOAP+PCA,J+22-ExoDtxn-PCA+GP}), though we take advantage of the improved computational efficiency of SOAPv4 and use four times the number stellar surface pixels used previously.

\section{From PCA to HSD: \texorpdfstring{\\}{ }Low-rank separable expansions for dynamic spectra}
\label{sec-decomp}

%-------------------------------------------------------------------------------
\subsection{Mitigating stellar activity using spectrum-based indicators}

Equation \ceq{Y-F0-lin-spar-time} formulated redshift (radial velocity) estimation as a nonlinear regression problem with the redshifts at all epochs, $z_k$, as the parameters of interest, and the parameters, $\spar$, describing the rest-frame dynamic spectrum, $\aspecem(\wob_{ik}, t_k;\spar)$, as a nuisance parameter.
Given the complex nature of stellar variability, modeling the rest-frame dynamic spectrum is challenging, likely requiring $\spar$ to be high-dimensional.
Until recently, astronomers adopted approaches for accounting for stellar activity that are simpler than this approach, aiming to adjust redshift estimates from traditional CCF or template fitting pipelines using one or a few summary statistics computed from the spectrum at each epoch (i.e., functionals of the spectrum at an epoch).
These summaries are called \emph{activity indicators} or \emph{activity proxies}, terminology borrowed from solar astronomy, where solar astronomers studying solar activity for its own sake have empirically identified several spectral features that are correlated with stellar activity.

\begin{figure}[t]
\centerline{\includegraphics[width=.95\textwidth]{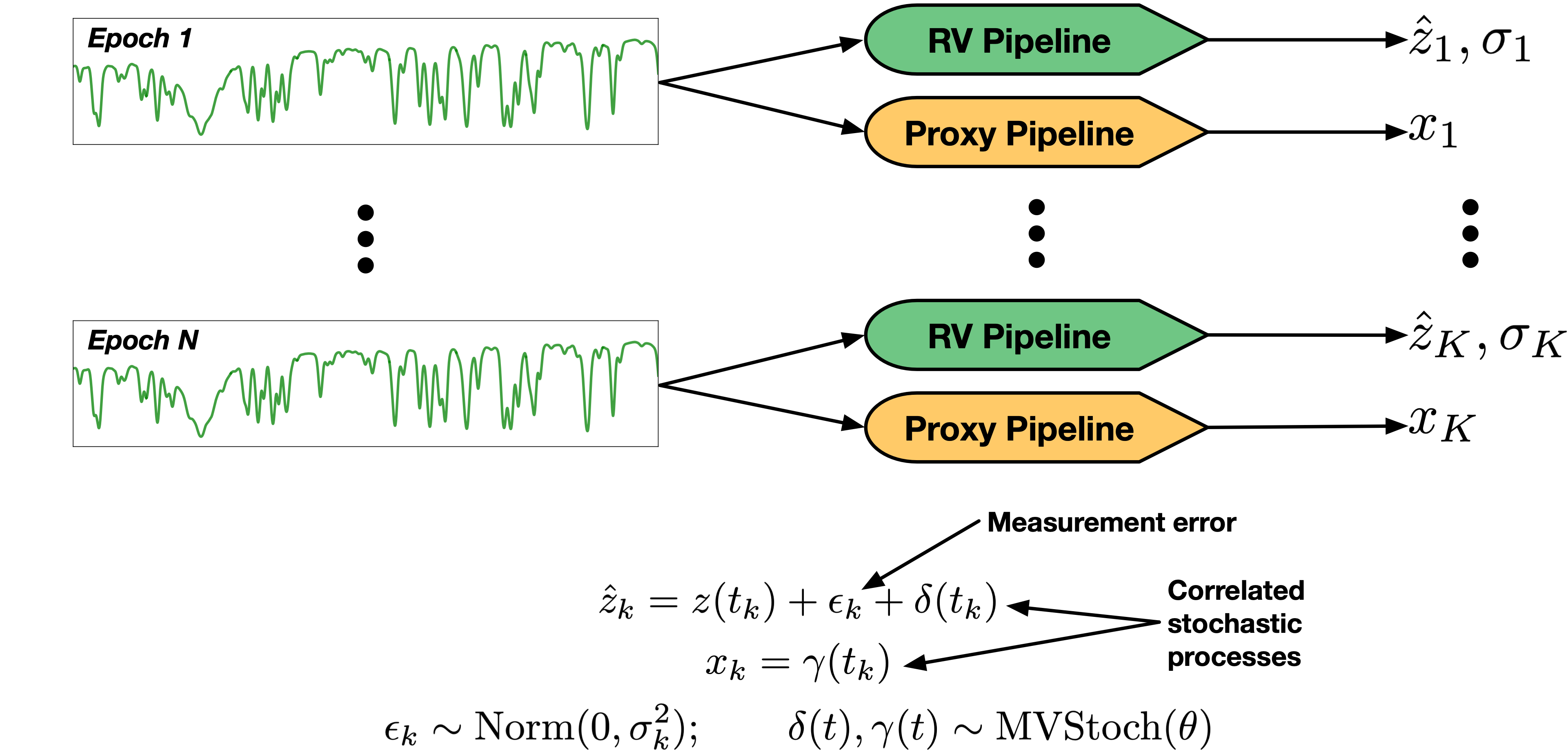}}
\caption{\small\baselineskip=11pt
Modeling stellar activity in RV pipeline outputs using spectrum-derived activity proxies as covariates/predictors.
$\operatorname{MVStoch}(\theta)$ denotes a multivariate stochastic process, i.e., a collection of two or more statistically dependent random functions of time, with parameters $\theta$.}
\label{RVPipeline-Proxies}
\end{figure}

\cfig{RVPipeline-Proxies} depicts this approach.
At each epoch, the RV pipeline produces a nominal redshift point estimate, modeled as the sum of the true redshift, a random measurement error contribution, $\epsilon_k$, and an activity-induced ``jitter'' contribution, $\delta(t_k)$:
\be
\hat z_k = z(t_k) + \epsilon_k + \delta(t_k).
\label{z-delta-proxy}
\ee
Separately from the redshift estimate, an indicator or proxy pipeline produces one or a few scalar summaries of the spectrum known to correlate with stellar activity, denoted here as a covariate, $x_k$.
Typical indicators include the intensities of one or more particular absorption lines correlated with activity, or measurements of the asymmetry of the shape of absorption lines (averaged over a chosen set of lines).
A relationship between the indicators and the jitter term is adopted, with the most sophisticated approaches positing a stochastic process model for the indicator time series that is correlated with the jitter time series.
These approaches have successfully mitigated the impacts of stellar activity on redshift to the level of a few tenths of a meter per second.
But this is still at least an order of magnitude above the level needed to enable detection of Earth-like planets around Sun-like stars.

%-------------------------------------------------------------------------------
\subsection{PCA-based stellar activity indicators}

In recent years various approaches that try to make more use of the full spectrum at each epoch have been developed, using various statistical and machine learning approaches.
None have yet achieved the performance needed to enable detection of exo-Earths.
Our work is motivated by a sophisticated extension of the indicator-based approach developed by a team of statisticians and astronomers working together as part of an astrostatistics program hosted by the Statistical and Applied Mathematical Sciences Institute (SAMSI).
Motivated by the work of \citet{D+17-SOAP+PCA} that used PCA to quantify the variability induced by starspots, \citet[J22]{J+22-ExoDtxn-PCA+GP} devised an approach like that depicted in \cfig{RVPipeline-Proxies}, with two innovations:
\begin{itemize}
\item Instead of using the conventional CCF or template approaches, the redshift point estimate is computed via linear regression in the manner suggested by \ceqn{Y-F0-lin-spar-time}, but with the spectrum derivative multiplying $z$ replaced with the time-averaged solar spectrum (i.e., not the instantaneous spectrum).
\item In place of conventional indicators, PCA was applied to a set of solar spectra to identify the top two principal components (forced to be orthogonal to the average wavelength derivative used for $\hat z_k$).
\end{itemize}
The use of the time-averaged spectrum derivative induces a jitter component into the redshift estimate.
The jitter is estimated by modeling its correlation with the PCA-based indicators by expressing $\hat z$ and the PCA scores as sums of latent Gaussian processes.
Each observable ($\hat z$ and the PCA scores) has its own GP, but all of them also have contributions by a shared GP, $X(t)$, and its first and second derivatives.%
\footnote{There is some astrophysical motivation for considering spot-induced activity effects to depend on a latent function and its derivative; this was the motivation for the dependence model, though no attempt was made to connect the model to spot astrophysics.}
Correlation is induced mainly by the shared $X(t)$-based terms, and indirectly by having the individual GPs share hyperparameters (all GPs used a quasi-periodic covariance function).
In simulation experiments, the method is able to detect planets with radial velocities approaching the 10~\cmps\ when stellar activity is due to a single large starspot.
J22 suggested several directions for future research: finding alternatives to PCA that might improve detection sensitivity, exploring other activity scenarios (multiple spots of different size, other types of activity), and, especially, extending the approach to apply to stars different from the Sun, by enabling star-specific discovery of indicators via PCA with observed stellar spectra.

%-------------------------------------------------------------------------------
\subsection{Low-rank separable expansions via HSD}

Our work is inspired by J22, particularly the idea of using empirical basis functions to model spectral variability.
Our departure point is to consider the rationale for use of PCA for this problem.

PCA targets analysis of fixed-dimension vector data (with components that are typically inhomogeneous, in the sense of having different physical units, thus requiring identification of a metric for the vector space).
When used for statistical inference, the vectors comprising the data are considered to be IID samples from a finite-dimension joint distribution.
These properties are poorly matched to the nature of EPRV spectral data.
A spectrum at a particular epoch comprises measurements of a continuous function, not of a finite-dimension vector.
Also, a dynamic spectrum evolves smoothly in time, so there is dependence between spectra measured at different times.
PCA ignores the functional nature of spectra, and the continuous evolution of a dynamic spectrum  over time.

The continuous function nature of the instantaneous spectrum can be accounted for by moving from PCA to \emph{functional} PCA (FPCA).
The theoretical foundation for FPCA is the Karhunen-Loeve expansion (KLE) for a stochastic process, expressing sample paths from a stochastic process as superpositions of orthogonal functions with random, uncorrelated coefficients.
The orthogonal functions are eigenfunctions of the covariance function for the process.
FPCA methods use finite data (measurements of an observed population of functions, each sampled over a finite set of times) to estimate the KLE for the underlying process describing the observed population.
There is a close mathematial connection between PCA and FPCA; the former represents vector data using eigenvectors of the sample covariance matrix; the latter represents functional data using eigenfunctions of an estimated covariance function.
Computationally, both approaches are typically implemented using singular value decomposition (SVD).
PCA often uses SVD as a convenient way to diagonalize the sample covariance matrix (it is more stable than a direct eigenvector/eigenvalue decomposition).
FPCA often uses SVD to estimate eigenfunctions by discretizing the covariance function and using SVD on the resulting matrix.
But while FPCA would treat a spectrum at a particular epoch as a function, in a way that PCA does not, it remains poorly suited to modeling dynamic spectra because it presumes spectra measured at different times represent independent sample paths (from a stochastic process over wavelength).

Despite these conceptual shortcomings of PCA and FPCA for modeling dynamic spectra, it is clear from prior work that PCA-based methods for EPRV data analysis are doing something useful.
Establishing a sound framework for this kind of modeling could potentially point toward improvements and generalizations.

We adopt the perspective that modeling observations of a dynamic spectrum should be approached as modeling a single realization of a 2-D function, rather than from the spectrum replication perspectives underlying PCA or FPCA.
The model building task then should be viewed, not statistically, but rather from the perspective of bivariate function approximation.
Of course, statistical considerations naturally arise once we consider estimation of a dynamic spectrum from noisy data, or joint estimation of an ensemble of related dynamic spectra (say, from similar stars), in which case replication over the population of 2-D functions could be addressed via ``next-generation'' functional data analysis.
Our focus here is on the model building, as a foundation for subsequent work on use of these models in statistical analysis.

The HSD is a tool from approximation theory that constructs an infinite separable basis function expansion for 2-D functions that is optimal in the sense of providing truncated (finite, low-rank) expansions that minimize integrated square error (ISE).
The \emph{symmetric} HSD underlies the covariance function expansion used in the KLE and FPCA; it is an infinite-dimensional generalization of the matrix diagonalization used to compute PCA.
In those applications, the diagonalization is applied to a covariance quantity---the covariance matrix of the vector data for PCA, and the covariance operator (or its discretization) for the KLE and FPCA.
The \emph{asymmetric} HSD enables optimal separable basis discovery for a 2-D function whose arguments index different spaces.
It is directly applicable to high-fidelity approximation of a single dynamic spectrum, a 2-D function over wavelength and time.
This motivates exploring the HSD as a foundation for statistical modeling of a single dynamic spectrum.
It provides an alternative to PCA that respects both the functional nature of a spectrum at a given epoch, and the functional nature of evolution of the spectrum with time.
% The HSD also provides a foundation for modeling populations of related dynamic spectra, in the manner of ``second generation'' FDA; this partly motivates our work, but we do not pursue this extension here.

We are specifically motivated by the use of the HSD in the \matlab\ \chebfun\ function approximation package, which uses the HSD to build efficient approximations of user-specified functions to \emph{machine precision}, which can then be manipulated for equation solving, integration, and other operations \citep{TT13-Chebfun2D,TT15-ContinuousMatrix}.

Let $\sbf_0(\lambda)$ denote the time-averaged spectrum,
\be
\sbf_0(\lambda) \equiv \frac{1}{T} \int \dif t\; \dspec(\lambda,t),
\label{time-avg}
\ee
where $T$ denotes the duration of the time interval being modeled.
Adopt a separable, low-rank model for the dynamic spectrum of the form
% \begin{equation}
% \dspec(\lambda,t) = \sbf_0(\lambda) + \sum_{\beta=1}^B c_\beta \sbf_\beta(\lambda)\tbf_\beta(t) + \delta(\lambda,t),
% \label{F-sep}
% \end{equation}
\be
\dspec(\lambda,t) = \sbf_0(\lambda) + \sum_{\beta=1}^B c_\beta \sbf_\beta(\lambda)\tbf_\beta(t) + \delta(\lambda,t),
\label{F-sep}
\ee
with \emph{spectral basis functions} (``speclets'') $\sbf_\beta(\lambda)$, and \emph{temporal basis functions} (``modulators'') $\tbf_\beta(t)$.
(Note that this is equivalent to an expansion starting with $\beta=0$, with $c_0\tbf_0(t) = 1$.)
The coefficients $c_\beta$ could be absorbed into the definitions of the $\sbf_\beta$ or $\tbf_\beta$ functions, but we leave them here to enable adopting normalization conventions for the speclets and modulators:
\be
\int \dif\lambda\, [\sbf(\lambda)]^2 = 1; \qquad
\int \dif t\, [\tbf(t)]^2 = 1.
\label{sq-norm}
\ee
Here $\delta(\lambda,t)$ denotes a residual component not accounted for by the expansion.
We will consider expansions accurate enough that $\delta(\lambda,t)$ is small compared to measurement errors, so we neglect it in the following (though we could consider modeling it with some simple stochastic process).

Define the \emph{centered dynamic spectrum} by $\cdspec(\lambda,t) \equiv \dspec(\lambda,t) - \sbf_0(\lambda)$.

When the full bivariate function $\dspec(\lambda,t)$ is available, the optimal choice for speclets and modulators follows from the asymmetric HSD; see \citet{S93-HistorySVD,S11-FredholmHilbertSchmidt} and \citet[\S\S~4.4, 4.6]{HE15-TheoreticalFDAOperators}.
The speclets and modulators are adjoint eigenfunctions solving coupled integral equations, with the $c_\beta$ quantities appearing as \emph{singular values} (akin to eigenvalues, and positive by convention):
\begin{align}
c_\beta \tbf_\beta(t) 
  &= \int \dif\lambda\; \cdspec(\lambda,t) \sbf_\beta(\lambda); \label{adj-eigen1}\\
c_\beta \sbf_\beta(\lambda) 
  &= \int \dif t\; \cdspec(\lambda,t) \tbf_\beta(t).
\label{adj-eigen2}
\end{align}
To find uncoupled equations for the basis functions, multiply \ceqn{adj-eigen1} by $\cdspec(\lambda',t)$ and integrate over time:
\be
c_\beta \int\dif t\; \cdspec(\lambda',t) \tbf_\beta(t) 
  = \int \dif t \int \dif\lambda\; \cdspec(\lambda',t) \cdspec(\lambda,t) \sbf_\beta(\lambda).
\ee
Using \ceqn{adj-eigen2}, the LHS can be written in terms of $\sbf_\beta(\lambda)$.
To simplify the result, define the \emph{symmetric spectral kernel},
\be
K(\lambda',\lambda) \equiv \int\dif t\; \cdspec(\lambda',t) \cdspec(\lambda,t).
\label{K-def}
\ee
Then we have integral eigenfunction equations for the speclets, with eigenvalues $c^2_\beta$:
\be
c^2_\beta \sbf_\beta(\lambda')
  = \int \dif\lambda\;  K(\lambda',\lambda) \sbf_\beta(\lambda).
\label{sbf-eigen}
\ee
A similar exercise for the modulators gives
\be
c^2_\beta \tbf_\beta(t')
  = \int \dif t\;  J(t',t) \tbf_\beta(t),
\label{tbf-eigen}
\ee
where $J(t',t)$ is a symmetric temporal kernel,
$J(t',t) \equiv \int\dif\lambda\; \cdspec(\lambda,t') \cdspec(\lambda,t)$.

%-------------------------------------------------------------------------------
\subsection{Computing the Hilbert-Schmidt decomposition}

In both statistical analysis of observed data, and computational approximation of functions (as in \chebfun), we need to solve the integral equations above given function values over a finite set of $(\lambda,t)$ sample points.
Computationally, we can approximate the integrals via quadrature rules that discretize in time and wavelength.
For wavelength integrals we adopt nodes and weights $(\lambda_l,v_l)$, so the wavelength integral of some function $g(\lambda)$ can be approximated as follows:
\be
\int\dif \lambda\; g(\lambda)  \approx  \sum_l v_l g(\lambda_l).
\label{l-quad}
\ee
For time integrals we adopt a rule with nodes and weights $(t_m,w_m)$, so the time integral of some function $h(t)$ can be approximated similarly,
\be
\int\dif t\; g(t)  \approx  \sum_m w_m h(t_m).
\label{t-quad}
\ee

We here explicitly describe integrating speclets; modulator integrals may be handled similarly.
The spectral kernel of \ceqn{K-def} can be computed as
\be
K_{l'l} \equiv K(\lambda_{l'},\lambda_l) 
\approx \sum_m w_m  \cdspec(\lambda_{l'},t_m) \cdspec(\lambda_l,t_m),
\label{K-quad}
\ee
with $l$ and $l'$ running over the same wavelength quadrature nodes.
The discretized version of the speclet equation, \ceqn{sbf-eigen}, is
\bal
c^2_\beta \sbf_\beta(\lambda_{l'})
  &\approx \sum_l  K_{l'l}\, v_l\, \sbf_\beta(\lambda_l) \nonumber\\
  &= \sum_l  \left[ \sum_m  w_m  \cdspec(\lambda_{l'},t_m) \cdspec(\lambda_l,t_m) \right] v_l\, \sbf_\beta(\lambda_l).
\label{sbf-quad}
\eal
This resembles an eigenvector equation in linear algebra, and we'd like to exploit techniques from computational linear algebra to solve for the speclets.
To do this, define a dynamic spectrum matrix as
\be
F_{lm} \equiv v_l^{1/2}\,w_m^{1/2} \cdspec(\lambda_l,t_m),
\label{F-def}
\ee
and a speclet vector as $\sbf_\beta = [v_1^{1/2}\sbf_\beta(\lambda_1), \ldots]^T$.
Then \ceqn{sbf-quad} takes the form
\be
c^2_\beta \sbf_\beta = F F^T \sbf_\beta,
\label{sbf-quad-m}
\ee
which now has the form of a matrix eigensystem equation.

For the modulators, we similarly define the modulator vectors by $\tbf_\beta = [w_1^{1/2}\tbf_\beta(t_1), \ldots]^T$.
The temporal kernel appearing in \ceqn{tbf-eigen} can be discretized as
\be
J_{m'm} \equiv J(t_{m'},t_m) 
\approx \sum_l v_l  \cdspec(\lambda_l,t_m') \cdspec(\lambda_l,t_m).
\label{J-quad}
\ee
The discretized version of \ceqn{tbf-eigen} is then
\be
c^2_\beta \tbf_\beta = F\trans F \tbf_\beta.
\label{tbf-quad-m}
\ee
Numerically, the best way to solve either eigensystem equation is to perform singular value decomposition of the $F$ matrix.
This need be done only once; the left and right singular vectors give the speclets and modulators.

% TODO: Check which are the left and right vectors.

We now can sketch an algorithm for computing discretized speclets and modulators:
\begin{enumerate}
\item Specify nodes and weights for the wavelength and time quadratures.
Since in general the times may not be a uniform grid, a (composite) trapezoid quadrature rule is a convenient algorithm and what we use here.
The wavelengths are finely sampled, and the trapezoid rule provides accurate wavelength quadrature.%
\footnote{A more sophisticated rule could be used as long as it is compatible with the wavelength grid we are using.
In particular, if the temporal grid is crude, an approach that imposes smoothness may be preferable to the trapezoid rule.
But as we note below, our primary interest in the speclets, so we have not pursued more sophisticated quadrature rules for the time dimension here.}
\item Compute the dynamic spectrum array, $\dspec(\lambda_l,t_m)$.
\item Compute the average spectrum, and use it to compute the centered dynamic spectrum array, $\cdspec(\lambda_l,t_m)$.
% \item Use that array directly to find PCA-based basis functions.
\item Compute the matrix $F$ given by \ceqn{F-def}.
\item Find the first several eigenvector-eigenvalue pairs of the square matrices $F F\trans$ and $F\trans F$ via SVD.
\item Recover the speclets from the eigenvectors by dividing the components by $v_l^{1/2}$, and the modulators by dividing their eigenvectors by $w_m^{1/2}$.
% \item Compare those speclets to the PCA basis functions.
\end{enumerate}

\section{Comparison of PCA and HSD}
\label{sec-comparison}

In this section we will compare the accuracy with which one can approximate a dynamic spectrum using spectral basis functions found via PCA, versus using the speclet-modular expansion provided by HSD.
PCA treats the spectral and temporal dimensions in fundamentally different ways; it provides basis function in wavelength, but considers time merely to be a replication index.
The asymmetric HSD is asymmetric in the sense of allowing the two arguments of a function to be in different spaces, but conceptually the two arguments are treated symmetrically, simply as real-valued function arguments.
Our use of it, expressed in \ceqn{F-sep}, treats wavelength and time somewhat asymmetrically, in that we have pulled out the time-averaged spectrum as a leading term.
This is neither necessary nor recommended for general use of the asymmetric HSD (e.g., it is not done in \chebfun).
We adopt this formulation here for two reasons.
First, we want to compare HSD to PCA.
This asymmetric treatment is a functional analog to PCA using centered data (spectra here), so adopting this formulation seeks HSD basis functions that will be as similar as possible to those from PCA.
(We will explore how breaking this asymmetry may enable futher improvement over PCA elsewhere.)
Second, and more importantly for our application, we noted at the start of the Introduction that a key feature of most  dynamic spectral data is \emph{strong asymmetry between the spectral and temporal sampling}.
It is often the case that such data is finely sampled in wavelength (thanks to high the resolution of the spectrograph), but more poorly sampled in time.
This is particularly true for exoplanet RV data; EPRV spectrographys typically measure spectra at hundreds of thousands of wavelengths.
But the observations of a target star requiring scheduling the sparse resource of telescope time, so there are typically only dozens to a few hundred epochs sampled.
As a result, we expect spectral basis functions to be recovered more accurately than temporal basis functions.

There is also astrophysical motivation for an asymmetric treatment.
At a given time, the spectrum of a star is a superposition of spectra from across its surface, reflecting the physics of the photosphere in quiet and active regions.
It is astrophysically likely that photospheric physics is very similar across a population of similar stars.
On the other hand, the temporal behavior will reflect conditions very specific to individual stars, such as the number and locations of starspots and other active regions, where (in latitude and longitude) the regions are at a given time, and the rotation period of the star.
So we expect that spectral basis functions should have similar structure across a stellar population, but temporal basis functions should vary significantly from star to star.
Combining this with the difference in sampling in the spectral and temporal dimensions motivates considering speclet discovery to be the primary goal for HSD, with its explicit use of modulators important mainly insofar as it accounts for irregular temporal sampling on speclet discovery.

In this section we report on ongoing work comparing use of PCA and HSD to describe dynamic spectra from SOAPv4.
We treat here a case that was the centerpiece for the previously cited work of \citet{D+17-SOAP+PCA} and \citet[J22]{J+22-ExoDtxn-PCA+GP}: a Sun-like star with an equatorial spot covering 1\% of its surface area (this represents a larger than typical spot for the Sun; our preliminary results with smaller spots, and spots away from the equator, produce similar results).
Though current EPRV instruments have over $10^5$ spectral pixels, for our simulations we focus on about 1/10th of the spectrum (the portion shown in the 2nd panel from the top in \cfig{fig:SOAPSpecZoom}, in the vicinity of prominent Calcium lines).
We explore the behavior of PCA and HSD over thousands of simulated temporal observing designs; using $\sim 10^4$ samples in wavelength enables exploring this many designs at reasonable computational expense.

The full SOAPv4 dynamic spectrum covers 27644 wavelength pixels, and spans one stellar rotation period, sampled finely in time at 496 evenly-spaced values of rotational phase (phase, mapped to [0,1] over one period, plays the role of time).
This does not represent realistic temporal sampling; some particularly well-observed targets may have that number of epochs of coverage, but the temporal sampling will be highly irregular.
The full dynamic spectrum is a starting point for the comparison.
The evenly-spaced samples imply nearly equal quadrature weights in time (the first and last trapezoid rule weights are $1/2$ times the interior weights).
So we expect the speclets from PCA and HSD to be nearly identical using the full dataset.

Our main focus is on how the approaches compare as the time sampling becomes coarser and more irregular.
So we consider subsamplings with a subset $K$ of the 496 phases, chosen randomly, replicating over the random sampling for each choice of $K$.
We consider the full-data HSD results as ground truth (the HSD basis functions are proved to be optimal in ISE); in any case, PCA and HSD results differ only slightly with the full data.

\begin{figure}[t]
\centerline{\includegraphics[width=.85\textwidth]{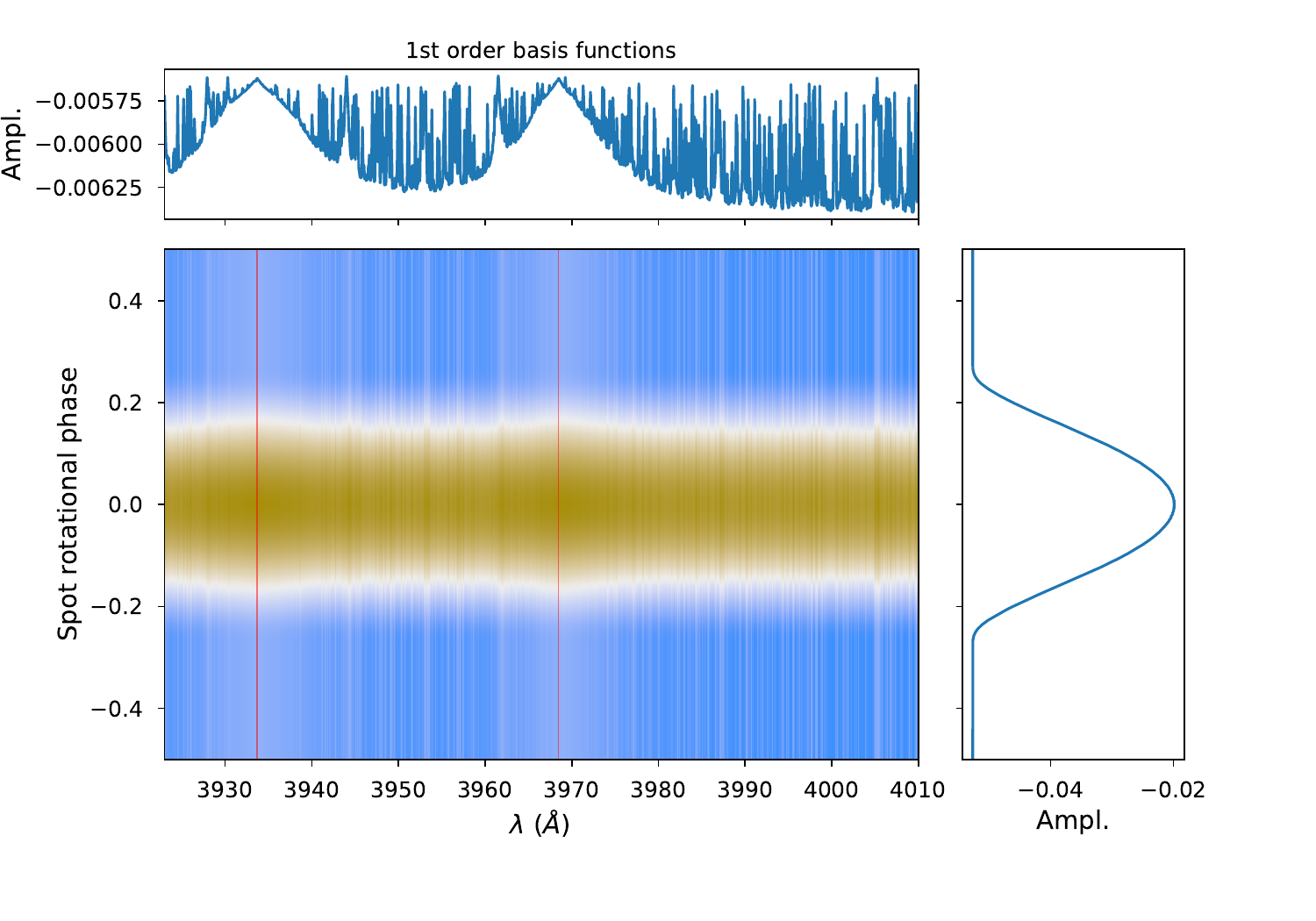}}
\caption{\small\baselineskip=11pt
First component of the HSD.
Top and right panels depict the 1st order speclet and modulator; image shows their net contribution to the difference between the dynamic spectrum and its time average.
The image color scale ranges from brown for lower values to blue for higher values.}
\label{SVD-1}
\end{figure}

\begin{figure}[t]
\centerline{\includegraphics[width=.85\textwidth]{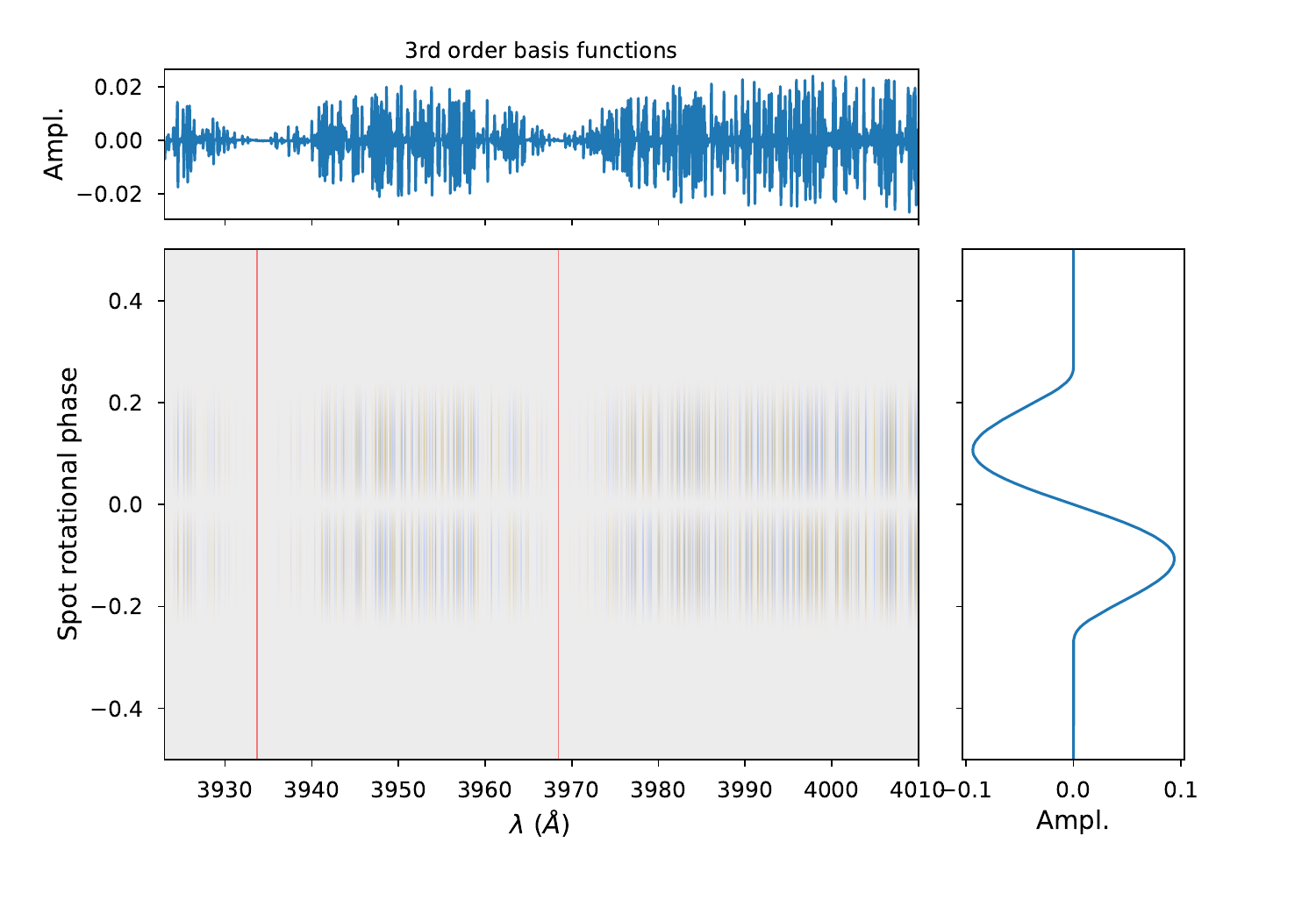}}
\centerline{\includegraphics[width=.85\textwidth]{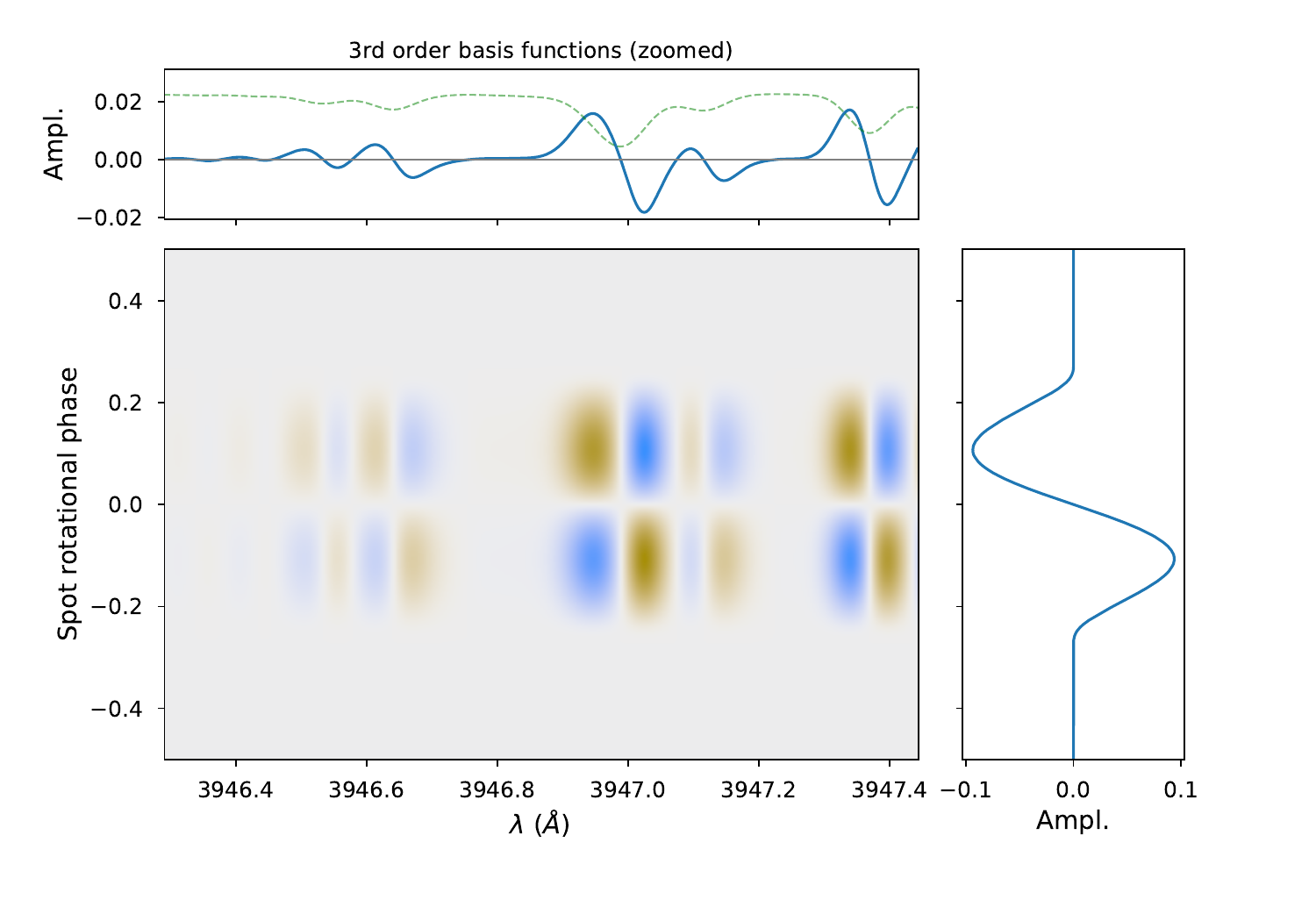}}
\caption{\small\baselineskip=11pt
As in \cfig{SVD-1}, but for the 3rd order speclet-modulator pair.
Lower plot shows a zoomed-in region of the upper plot.}
\label{SVD-3}
\end{figure}

\cfig{SVD-1} and \cfig{SVD-3} depict two terms of the HSD, as illustrations of how the HSD decomposes the dynamic spectrum.
\cfig{SVD-1} shows the first order term (after the time-average spectrum term).
It appears to capture a kind of average impact of the spot on the spectrum, in the sense of behavior that smoothly subtracts from the spectrum as the spot comes into view (middle phase values, near 0).
\cfig{SVD-3} shows the third order term, which has a more interesting modulation, changing sign as the spot moves toward (negative phases) and away (positive phases) from the observer.
The zoomed version (bottom plot) reveals that the net effect of this term will be to shift the locations of features in the spectrum.
This kind of behavior can mimic a periodic Doppler shift, so it is important to model it as accurately as possible.

\begin{figure}[t]
\centerline{\includegraphics[width=.85\textwidth]{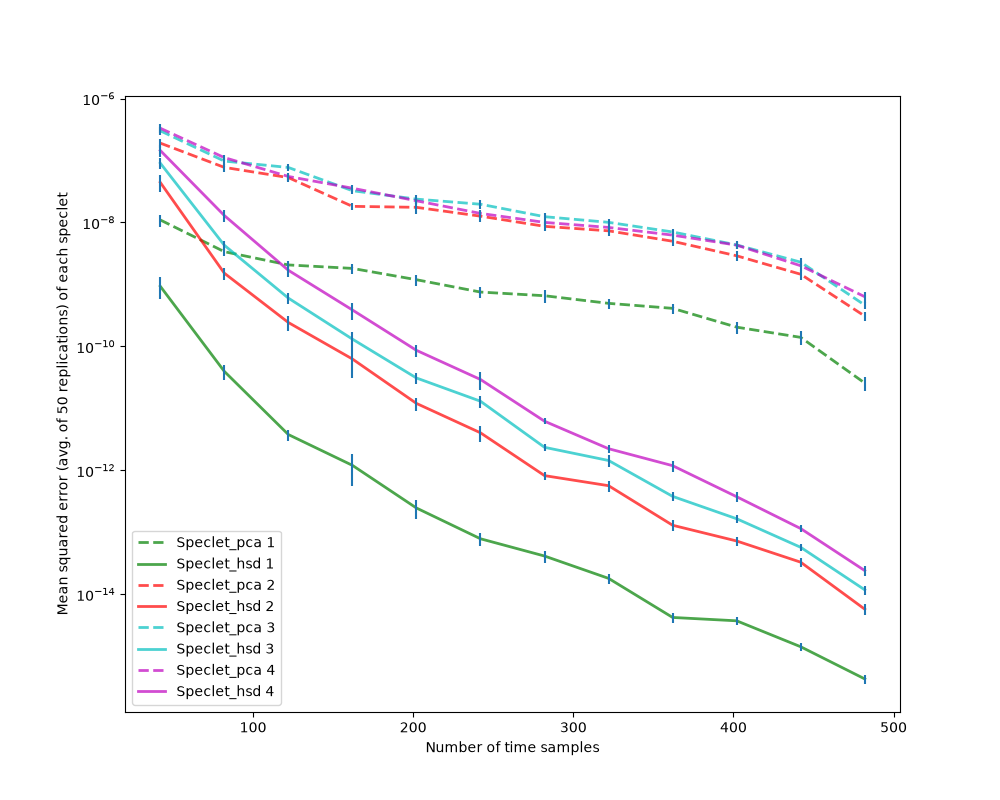}}
\caption{\small\baselineskip=11pt
Mean squared error (over wavelength) between recovered speclets and the full-data speclets, for recovery by PCA (dashed curves) and HSD (solid curves), for the 1st through 4th leading speclets, as a function of the number of sample times included in 50 random temporal subsamples of the full data.
Points and error bars show the mean and standard deviation of the MSE across the 50 realizations.}
\label{speclet-MSE}
\end{figure}

\cfig{speclet-MSE} shows how well speclets are recovered by the two approaches as a function of the number of time samples used, chosen randomly from the 496 samples in the full dataset (for 50 realizations).
Recovery accuracy is measured by mean squared error (MSE, i.e., squared error averaged over the wavelength or time sample points, with the mean and standard deviation over the 50 realizations plotted as points and error bars).
Results are shown for the four leading-order speclets.
For each order (color), the solid curve shows the MSE for the HSD recovery, and the dashed curve (of the same color) shows the MSE for the PCA recovery, computed.
For every order, the HSD MSE is significantly smaller than the PCA MSE.
The HSD speclet MSE gets vanishingly small (approaching machine precision) as the number of time samples approaches the full size of the simulated dynamic spectrum, since the HSD speclet is used for ground truth (Schmidt proved it is optimal with respect to squared error).
It remains significantly smaller than the PCA MSE even when only a small fraction of the 496 temporal samples are used (randomly chosen and thus irregularly spaced in time).

% TODO: We probably should use ISE, not MSE.

\begin{figure}[t]
\centerline{\includegraphics[width=.85\textwidth]{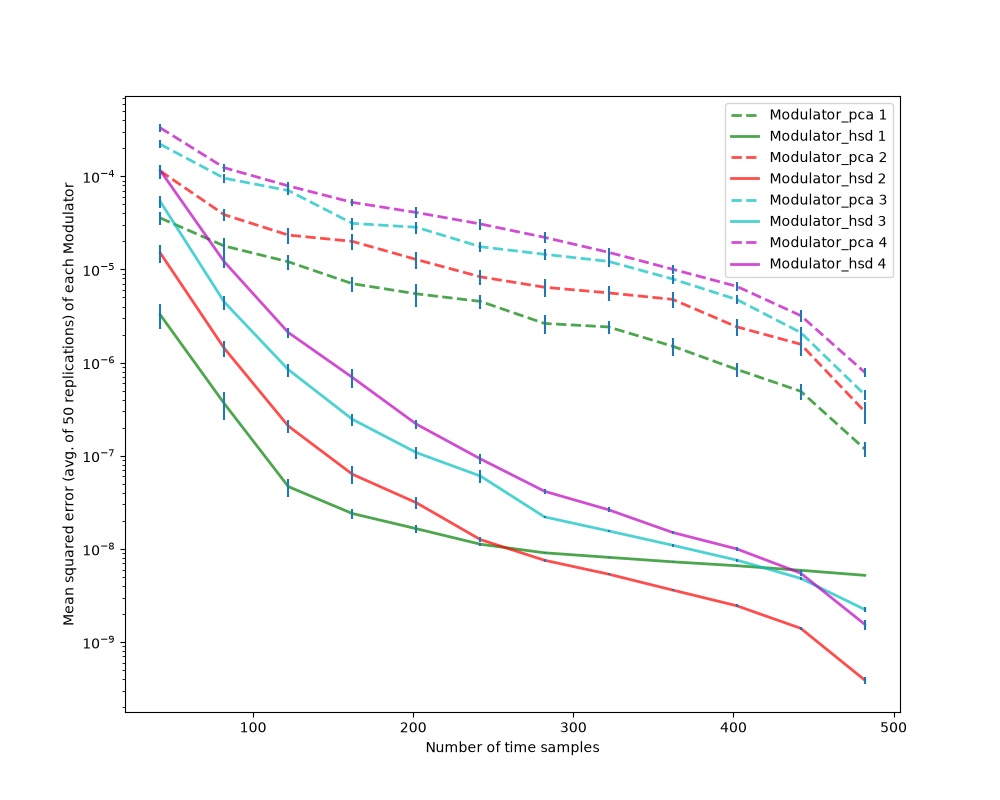}}
\caption{\small\baselineskip=11pt
Mean squared error (over time) between recovered modulators and the full-data modulators, for recovery by PCA (dashed curves) and HSD (solid curves), for the 1st through 4th leading modulators, as a function of the number of sample times included in 50 random temporal subsamples of the full data.
Points and error bars show the mean and standard deviation of the MSE across the 50 realizations.}
\label{modulator-MSE}
\end{figure}

\cfig{modulator-MSE} shows results from a similar calculation for the modulators, of less interest than the speclets, but presented for completeness. HSD continues to show better recovery.

\begin{figure}[t]
\centerline{\includegraphics[width=.85\textwidth]{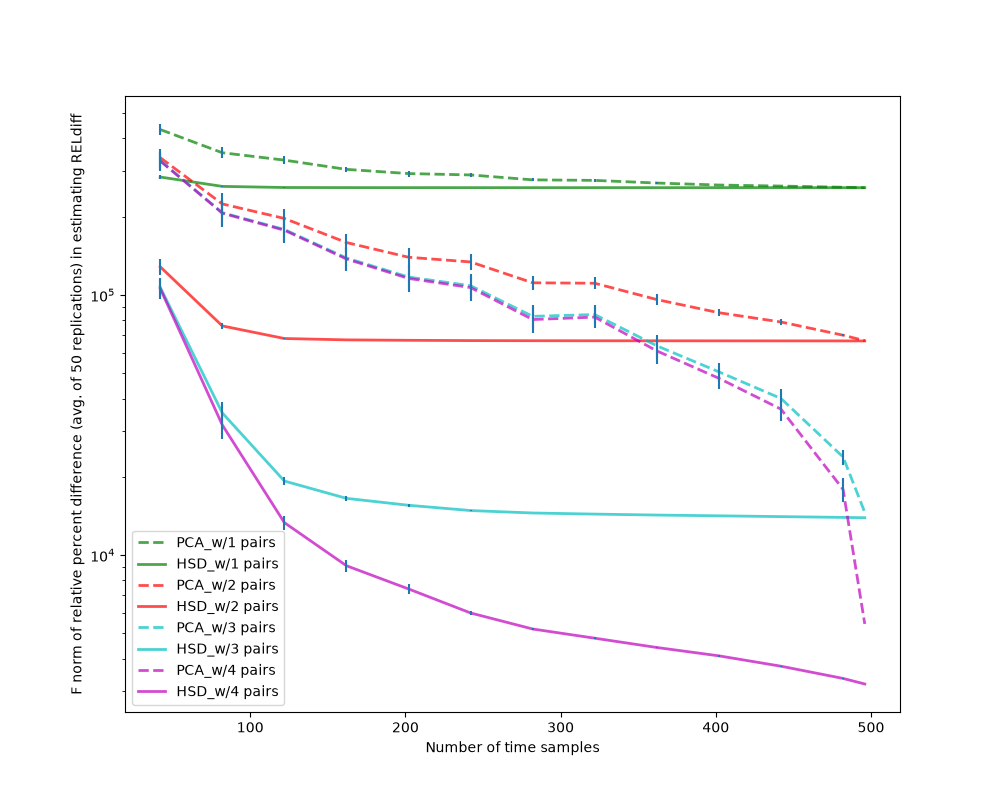}}
\caption{\small\baselineskip=11pt
Mean Frobenius norm (F norm) between the reconstructed dynamic spectrum and the true dynamic spectrum, for recovery by PCA (dashed curves) and HSD (solid curves), using from 1 to 4 speclet-modulator basis function pairs, as a function of the number of sample times included in 50 random temporal subsamples of the full data.}
\label{dynspec_Fnorm}
\end{figure}

\cfig{dynspec_Fnorm} shows the average reconstruction error for the dynamic spectrum, measured using the Frobenius norm, using HSD and PCA.
For the full dataset (496 time samples), both approaches reconstruct the dynamic spectrum with equal accuracy, though the accuracy is poor if only the leading order speclet/modulator pair is used.
For reconstructions of higher order, as the number of randomly chosen time samples decreases (with increasing irregularity in the time sampling), the reconstruction error using PCA grows quickly, while the error using HSD is significantly lower.
The HSD error is about an order of magnitude lower than the PCA error if three or four speclet/modulator pairs are used, until the most coarse subsampling is reached (42 time samples), at which point the HSD error is still about three times smaller than the PCA error.

% TODO: Label images by wavelength and phase

\begin{figure}[t]
\centerline{\includegraphics[width=.85\textwidth]{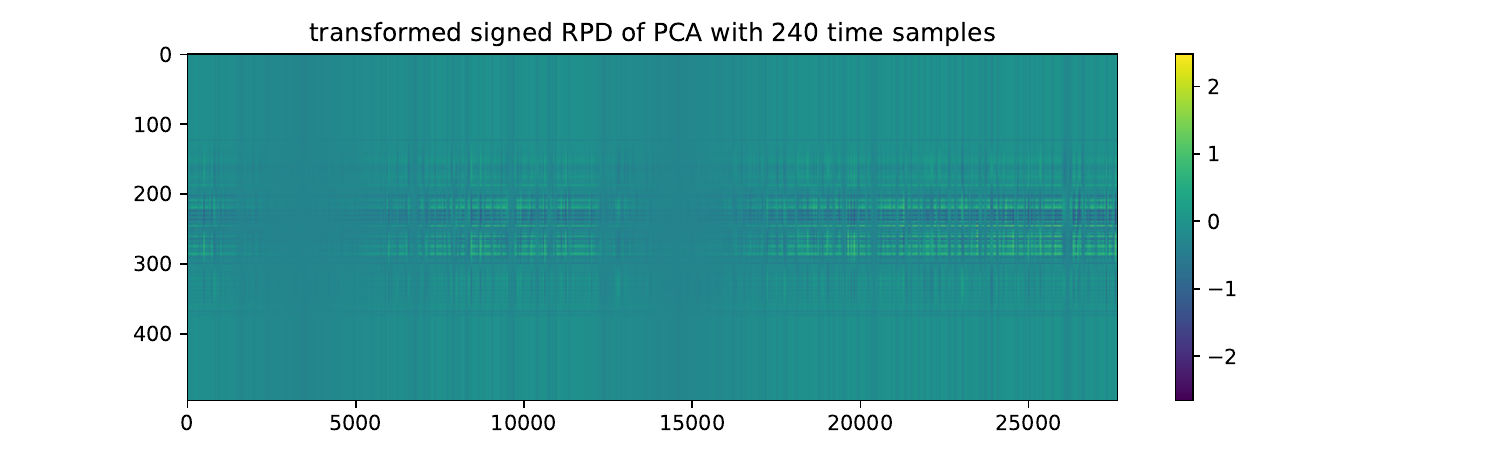}}
\centerline{\includegraphics[width=.85\textwidth]{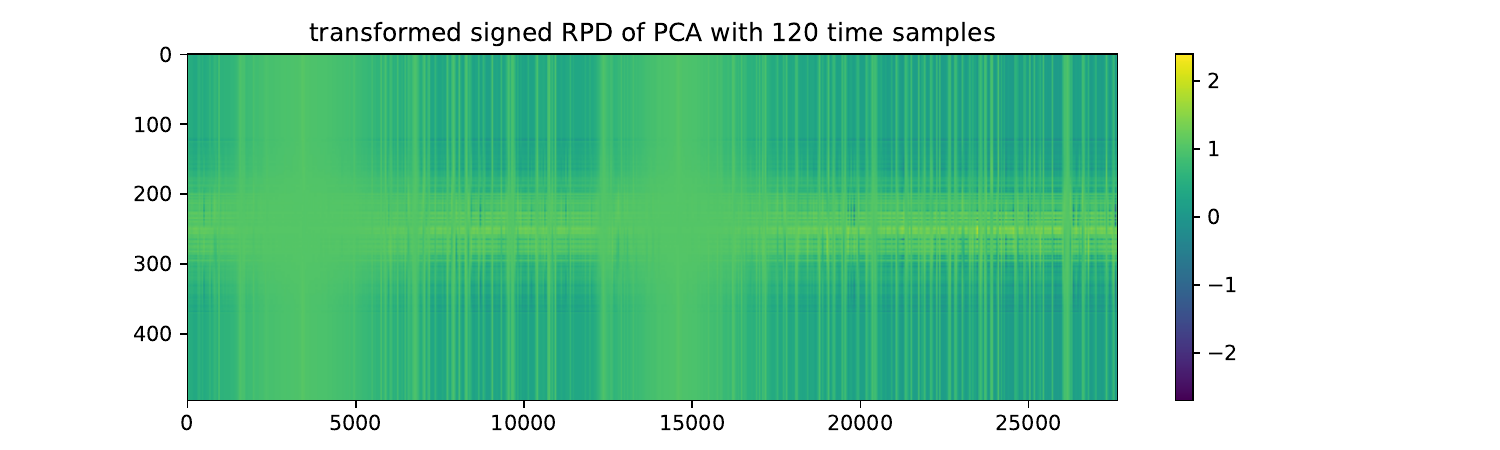}}
\caption{\small\baselineskip=11pt
Error in the PCA recovery for the dynamic spectrum, shown as a relative percent difference (RPD) image between the reconstructed spectrum and the actual dynamic spectrum, using 240 (top panel) and 120 (bottom panel) subsamples of the 496 time samples in the full simulated dynamic spectrum.
A nonlinear transformation was used to enhance the visibility of the images.}
\label{RPD-trans}
\end{figure}

\cfig{RPD-trans} shows images of the (signed) relative percent difference (RPD) between the PCA reconstruction of the dynamic spectrum and the actual dynamic spectrum (wavelength horizontal by pixel number, time vertical by phase pixel number).
These plots show the functional nature of the PCA reconstruction error; it is largest as a function of phase in the middle phase regions, corresponding to times when the spot is visible, and it is largest as a function of wavelength inside absorption lines---the most important part of the spectrum to recover accurately.

These results demonstrate that HSD not only offers a more sound conceptual framework for empirical basis function discovery for EPRV dynamic spectra than does PCA, but that the new conceptual framework leads to revised calculations that significantly improve the accuracy of models of a fixed order.

% \clearpage

\section{Summary and future directions}
\label{sec-summary}

Truncated expansions are a widely used tool for both statistical modeling and approximation of functions.
For univariate functions, they appear in the form of finite expansions using a pre-defined basis, such as Fourier or orthogonal polynomial bases, or using adaptive or empirical basis functions, as is done by FPCA in statistical settings, or proper orthogonal decomposition (POD) in deterministic function approximation settings.
Both FPCA and POD typically use SVD in their computational implementations.
For multivariate functions, a natural generalization of the idea is the use of low-rank separable expansions.
For bivariate functions, the HSD can be used to construct orthogonal basis functions in each dimension that produce optimal low-rank representations.
Their discretization leads to a computational implementation that uses SVD, but used differently than how it is used for PCA.
The appearance of SVD across these approaches invites confusion regarding matching a method to a problem.

The main message of this work is that since dynamic spectra are bivariate functions, model construction should explicitly recognize their bivariate nature.
Empirical basis function construction should recognize the functional nature of a spectrum observed at a particular epoch (an aspect ignored by PCA), and the functional nature of the time evolution (an aspect ignored by FPCA and the Karhunen-Loeve expansion).
For modeling dynamic spectra of a Sun-like star with a single starspot, we showed that use of HSD produces basis functions that can capture spectral variability more accurately than using basis functions computed via PCA (which treats the collection of observed spectra as an ensemble of IID vector samples).

EPRV dynamic spectra are finely sampled in wavelength but sparsely and irregularly sampled in time.
Both the nature of the temporal sampling, and astrophysical considerations, suggest that temporal basis functions (modulators) will be case-specific and will likely require nontrivial statistical modeling to account for the sparsity, and noise (e.g., using multivariate GP modeling as described by \citet{J+22-ExoDtxn-PCA+GP}).
But our calculations show that even basic consideration of temporal sampling, at the somewhat simple level used by our HSD implementation (which implicitly interpolates linearly in time, via use of the trapezoid rule for discretization), produces spectral basis functions (speclets) that can be significantly more accurate than those constructed by PCA (i.e., ignoring the temporal design).

An immediate question is how well a spectral basis set can describe situations that differ from the setting they were optimized for.
We have begun exploring this for HSD modeling of starspot activity by using SOAPv4 to simulate activity with diverse active regions.
Our current work considers single spots, but with different sizes and stellar latitudes than the single equatorial spot used for the calculations of \S~\ref{sec-comparison}.
We are comparing the speclets found across the varied settings, and also evaluating how well speclets from, say, the equatorial spot calculations of \S~\ref{sec-comparison}, can recover dynamic spectra from other spot geometries.
Our initial results suggest that the speclets are very similar, with the modulators being the main difference across the settings---approximate universality of speclets.
Future work will consider variability from faculae, which could produce speclets that differ significantly from those derived from starspot variability.
If that is the case, analyses of EPRV spectral data using separable expansions will need to include a library of speclets covering the relevant types of stellar activity.

Basis functions from HSD are optimal in the sense of integrated squared error over the dynamic spectrum.
This is a natural criterion from a function approximation perspective, with the virtue of being amenable to theoretical analysis (e.g., Schmidt's demonstration of ISE optimality).
But the ultimate goal of the model is to enable accurate redshift estimation, e.g., via regression using \ceqn{Y-F0-lin-spar-time}.
In redshift estimation the rest-frame dynamic spectrum is a nuisance parameter.
This motivates considering metrics that specifically target redshift estimation.
In particular, redshift enters \ceqn{Y-F0-lin-spar-time} as the multiplier of the \emph{wavelength derivative} of the dynamic spectrum.
This suggests using derivative recovery metrics, as a supplement to the ISE metric that HSD targets, or possibly as the basis of an alternative decomposition.

\citet{J+22-ExoDtxn-PCA+GP} suggested considering alternatives to PCA for constructing spectral basis functions, such as manifold learning techniques from machine learning.
Our study indicates such alternatives should recognize the bivariate nature of the problem.
An intriguing possibility is to exploit nonnegativity of spectra; requiring basis functions to be nonnegative may produce more interpretable speclets, and perhaps expand the universality of speclets across sources of variability.
In settings where PCA has traditionally been used, nonnegative matrix factorization (NMF) is sometimes advocated as an alternative to PCA; empirically, the constrained basis functions are often more interpretable than PCA basis functions \citep{LS99-NMF}.
There are generalizations of the NMF idea to functional and multivariate data that could be considered as alternatives to HSD for dynamic spectrum representation.

Our focus has been on model construction; deployment for statistical modeling of actual EPRV data is a natural next step.
EPRV observers share data comprising redshift estimates (with uncertainties), and often spectral summaries (indicators), across epochs; but to our knowledge it is not yet common for full spectra to be shared for every epoch.
One possibility is for instrument teams to supplement the shared indicators with speclet-derived indicators (functionals of the observed spectra), to enable analysts to develop statistical models that make use of projections onto optimal speclets.
Because the adjustments made to ``normalize'' spectra are instrument-specific, it may be the case that each observing team will have to tailor speclet calculation to the instrument pipeline.

Finally, we here target analysis of Sun-like stars, because we have access to dynamic spectra of the Sun that are free of planet-induced Doppler shifts.
We would like to deploy HSD or similar approaches for analyzing data from target stars of diverse stellar types.
SOAPv4 supports use of spectra computed from sophisticated theoretical codes capable of modeling photospheres of diverse types of stars (e.g., the PHOENIX code, \citet{H+13-PHOENIX}).
This capability could be used to derive optimal speclets as a function of stellar type.
But as sophisticated as modern theoretical calculations are, they do not achieve perfect fidelity, and they can only roughly model stellar activity (e.g., they typically do not model magnetohydrodynamic phenomena underlying activity).
An alternative approach would be to fit dynamic spectrum observations of an EPRV target star jointly for separable basis functions and the orbit of a potential planet (or planets).
It is not obvious that this is feasible for observations of a single target (e.g., there could be identifiability issues, particularly regarding sources of quasiperiodic stellar activity).
More ambitiously, this kind of joint fitting could be attempted across a \emph{collection} of targets (of similar stellar type), e.g., in a hierarchical Bayesian functional data analysis framework that would enable borrowing strength across targets to help disentangle variability signals (expected to have some similarity across targets) from exoplanet signals (which should be target-specific).

%===============================================================================
%===============================================================================

%% For this sample we use BibTeX plus aasjournalv7.bst to generate the
%% the bibliography.
%% Use longbib in the \documentclass specification to show titles.
%% To get the citations to show in the compiled file do the following:
%%
%% pdflatex <basename>.tex
%% bibtext <basename>
%% pdflatex <basename>.tex
%% pdflatex <basename>.tex

\newpage

\bibliography{EPRV-HSDReport}{}
\bibliographystyle{aasjournalv7.1}

\end{document}